\documentclass[journal]{new-aiaa}
\usepackage[utf8]{inputenc}
\usepackage{subfigure,subfigmat}
\usepackage{multicol}
\usepackage{verbatim}
\usepackage{graphicx}
\usepackage{array}
\usepackage{amsmath}
\usepackage[version=4]{mhchem}
\usepackage{siunitx}
\usepackage{color}
\usepackage{soul}
\usepackage{nomencl}
\usepackage{booktabs}
\usepackage{enumitem}
\usepackage{psfrag}
\usepackage{longtable,tabularx}
\usepackage{appendix}
\usepackage[colorinlistoftodos]{todonotes}

\title{Full trajectory optimizing operator inference for reduced-order modeling using differentiable programming}

\author{Surya Chakrabarti \footnote{Ph.D. Student, chakrabarti.23@osu.edu}, Datta Gaitonde\footnote{John Glenn Professor, gaitonde.3@osu.edu}}
\affil{The Ohio State University, Mechanical and Aerospace Engineering Department, Columbus, Ohio, 43210}
\author{Arvind T. Mohan\footnote{Scientist, Computational Physics and Methods Group,}, Daniel Livescu\footnote{Scientist, Computational Physics and Methods Group,}}
\affil{Los Alamos National Laboratory, Los Alamos, New Mexico 87545}

\usepackage{xcolor}
\usepackage{soul}

\begin{document}
\maketitle
\begin{abstract}
Accurate and computationally inexpensive Reduced Order Models~(ROMs) capable of forecasting turbulent flows can facilitate rapid design iterations and thus prove critical for predictive control strategies in engineering problems.
Galerkin projection based Reduced Order Models~(GP-ROMs), derived by projecting the Navier-Stokes equations on a truncated Proper Orthogonal Decomposition~(POD) basis, are popular because of their low computational costs and theoretical foundations in the governing fluid flow equations. 
However, the accuracy of traditional GP-ROMs degrades over long time prediction horizons. 
While calibration can somewhat mitigate this issue, the procedure is cumbersome  and requires insight into the full order solver. 
To address these issues, we extend the recently proposed \textit{Neural Galerkin Projection (NeuralGP)} data driven framework to compressibility-dominated transonic flow, considering a prototypical problem of a buffeting NACA0012 airfoil governed by the full Navier-Stokes equations. 
The algorithm maintains the form of the ROM-ODE obtained from the Galerkin projection;  however coefficients are learned directly from the data using gradient descent facilitated by differentiable programming. 
This effectively blends the strengths of the physics driven GP-ROM and purely data driven neural network-based techniques, resulting in a computationally cheaper model that is easier to interpret. 
While both NeuralGP and traditional GP-ROMs yield accurate forecasts over short time prediction horizons, the former accurately tracks the reference POD coefficients over longer prediction horizons, in time ranges where the calibrated GP-ROM displays the tendency to diverge in amplitude and frequency. 
We show that the distinct long-term behavior of the blended NeuralGP method minimizes a more rigorous full trajectory error norm compared to a linearized error definition optimized by the calibration procedure. 
We also find that while both procedures stabilize the ROM by displacing the eigenvalues of the linear dynamics matrix of the ROM-ODE to the complex left half-plane, the NeuralGP algorithm adds more dissipation to the trailing POD modes resulting in its better long-term performance. 
The results presented highlight the superior accuracy of the NeuralGP technique compared to the traditional calibrated GP-ROM method.

\end{abstract}

\section{Introduction}
\label{sec:intro}

Incorporating a robust, active feedback control for turbulent flows is a promising research problem with the potential to greatly improve the design of flight vehicles and more efficient energy systems~\citep{gad2001flow,bons2005designing,samimy2007active,wei2006noise}.  
The goal of controlling chaotic and nonlinear systems like turbulent flow is greatly enabled by the capability to making predictions for several time instants into the horizon in a fast, inexpensive and accurate manner. 
High fidelity simulations of fluid flows, while accurate, can be prohibitively expensive for efficient forecasting and prediction-based control techniques. 
This is particularly the case for 3D high Reynolds number turbulent flows, where the number of degrees of freedom can exceed billions, and requires substantial resources on supercomputers.

The increased push towards developing control strategies for turbulent flows has led to some focus on the development of Reduced Order Models~(ROM) that retain the accuracy of  high fidelity simulations while reducing the computational resources needed for predictions made over time horizons spanning several non-dimensional flow times~\cite{rowley2004model,noack2011reduced}. 
One of the more popular strategies uses Galerkin projection of the Navier-Stokes equations onto a set of truncated orthogonal basis vectors, which are usually obtained from a Proper Orthogonal Decomposition~(POD) of the flow fluctuations. 
The resulting projected equations may be solved at relatively little computational cost, while capturing the dominant physics of the flow as discussed in Refs.~\citep{rempfer2000low,rowley2004model,lorenzi2016pod, benner2015survey,chinesta2016model}. 
Several of these exploit assumptions such as  isentropic flow which can constrain parameters where they may be used, such as moderate Mach numbers.
ROMs for compressible flow problems governed by the full Navier-Stokes equations, while possible~\cite{bourguet2011reduced}, have not been as widely explored.

Althought their theoretical formulation is directly predicated on the Navier Stokes equations, traditional GP-ROMs suffer from structural instabilities~\cite{rowley2004model,callaham2021role} that severely limit their prediction capabilities and result in model divergence over longer time horizons. 
This has prompted the development of several optimization techniques aimed at reducing prediction errors and stabilizing the ROM solutions.
Examples include non-linear constrained optimization~\cite{bergmann2004optimisation}, least squares minimization~\cite{couplet2005calibrated} and Tikhonov regularization~\cite{cordier2010calibration}. 
\citet{cordier2010calibration} compared several calibration methodologies and concluded that the technique based on Tikhonov regularization outperformed all other methods. 
Despite the application of state-of-the-art calibration techniques however, long-term predictions from GP-based ROMs remain unreliable~\cite{mohan2021learning}, necessitating newer methods for improved prediction performance. Furthermore, the construction of a GP-ROM is an intrusive process, requiring knowledge of the governing equations and the discretization schemes employed in the full order model~\cite{peherstorfer2016data}. 
Additionally, the introduction of new problems with additional physics, such as reacting flows, requires a re-derivation of the ROM Ordinary Differential Equation~(ODE), further limiting the generalizability of such methods.

Recent studies have leveraged deep learning-based approaches, such as neural networks (NNs), to create data-driven alternatives to GP-ROMs with the goal of improving long-term predictions from ROMs, while also avoiding the intrusiveness of traditional ROM methodologies.  
Several variants of NNs exist, with the Long Short-Term Memory~(LSTM) architecture~\citep{hochreiter1997long} being popular in turbulence~\citep{mohan2018deep,ahmed2020long,deng2019time,maulik2020non} owing to its strengths in time-series prediction. 
Although the literature reports that NNs have considerable promise, they face important limitations to practical use because of their black-box nature, lack of interpretability and the requirement of relatively large amounts of training data. 
To avoid these problems associated with NNs for ROMs, \citet{peherstorfer2016data} proposed a non-intrusive operator inference method which retains the form of the ROM-ODE obtained from the Galerkin Projection, but seeks to estimate its coefficients directly using training data. 
While a prior knowledge of the form of the ROM-ODE is required for implementing this method, it can be easily obtained owing to the properties of the Galerkin projection. 
For a full order model given by a Partial Differential Equation~(PDE) with a polynomial non-linearity, the Galerkin projection onto a reduced basis preserves the structure of the equations and results in a ROM-ODE with a similar form as the original full order PDE~\cite{ghattas2021learning}.

The operator inference method, however, relies on a linear optimization technique that operates on training data comprising discretely sampled points on the trajectories of the projected reduced state and as such, does not account for the errors incurred over the full predicted trajectory obtained by integrating the ROM-ODE. Recently, \citet{mohan2021learning} proposed a hybrid learning approach called \textit{Neural Galerkin Projection (NeuralGP)} that uses deep learning to derive the coefficients for a ROM-ODE of a known form.
The method uses \textit{Differentiable Programming}~(henceforth referred to as \textit{DiffProg}) - a programming paradigm that enables software to perform full differentiation via automatic differentiation~\citep{griewank1989automatic,bartholomew2000automatic,baydin2018automatic}. 
This facilitates calculation of gradients of the solutions from generalized ODE/PDE solvers with respect to the ODE coefficients, allowing for their optimization using gradient-descent based methods. 
We tested this algorithm for an isentropic flow, and while it showed considerable promise yielding superior results compared to an isentropic GP-ROM, the simplifying assumptions of constant entropy significantly limits practical applications. 

The present paper thus seeks to extend the \textit{NeuralGP} method to a high-speed compressible flow problem governed by the full Navier-Stokes equations and compare the resulting ROM-ODEs and their predictive capability against the more widely studied GP-ROMs.
As a test case, we examine transonic buffet flow over a NACA0012 airfoil, representing a well-examined aerospace application often  used as a test bed for feedback control~\cite{caruana2005buffet,ren2020adaptive} and ROM~\cite{bourguet2011reduced} studies. 
We summarize the numerical method used for the simulation in Sec.~\ref{sec:methodology} while a detailed description of the flowfield features are provided in Sec.~\ref{sec:flow_feat}. 
Two different ROMs for the flowfield are then constructed using the GP-ROM and \textit{NeuralGP}, respectively. 
We outline a description of the procedure used to construct the ROM along with key details of the mathematical derivation of each method in Sec.~\ref{sec:ROM_method}. 
We evaluate the short and long-term predictions of the NeuralGP in Sec.~\ref{sec:results} and identify benefits relative to prior approaches
In Sec.~\ref{sec:sysDynAnalysis}, 
we analyze the ROM-ODEs from a dynamical systems perspective, with the goal of explaining factors that contribute to improvement in prediction performance. 
Finally, we summarize our contributions in Sec.~\ref{sec:conclusions}.


\section{Numerical Method}
\label{sec:methodology}

The non-dimensionalized compressible Navier-Stokes equations cast in the strong conservation form are solved in a curvilinear coordinate $(\xi, \eta)$ frame with $J$ being the Jacobian of the curvilinear transformation given by $\bigg(J=\frac{\partial (\xi,\eta,\tau)}{\partial (x,y,t)}\bigg)$~\cite{anderson2016computational}.
\begin{equation}
  \frac{\partial}{\partial \tau}\bigg(\frac{\overrightarrow{Q}}{J}\bigg)+\frac{\partial \hat{F}}{\partial \xi}+\frac{\partial \hat{G}}{\partial \eta}=\frac{1}{Re}\bigg[\frac{\partial \hat{F_{\nu}}}{\partial \xi}+\frac{\partial \hat{G_{\nu}}}{\partial \eta}\bigg]
  \label{eq:NS_equn}
\end{equation}
where  $\overrightarrow{Q}$ refers to a vector comprised of the conserved variables. 
Thus, $\overrightarrow{Q}=[\rho,\rho u,\rho v,\rho  E]^T$. The vectors $\hat{F}$,  and  $\hat{G}$ correspond to the inviscid fluxes and the vectors $\hat{F_{\nu}},\hat{G_{\nu}}$ constitute the viscous fluxes. 
Thus,
\begin{equation}
  F=\left[\begin{array}{c}
      \rho \hat{U}\\
      \rho u \hat{U}+\xi_x p\\
      \rho v \hat{U}+\xi_y p\\
      
      (\rho E + p)\hat{U}\\
    \end{array}\right]
  \quad and \quad
  F_{\nu}=\left[\begin{array}{c}
      0\\
      \xi_{x_i} \sigma_{i1}\\
      \xi_{x_i} \sigma_{i1}\\
      \xi_{x_i}  (u_j\sigma_{ij}-\Theta_i)\\
    \end{array}\right]
\end{equation}
The contravariant velocity component is given by $\hat{U}=\xi_x u+\xi_y v$ and the specific energy density is
\begin{equation}
  E=\frac{T}{\gamma (\gamma -1)M^2_\infty} + \frac{1}{2}(u^2+v^2)
  \label{eq:sp_energy}
\end{equation}
The deviatoric stress~($\sigma_{ij}$) and heat flux vector~($\Theta_i$) are given by:
\begin{equation}
  \sigma_{ij}=\mu\bigg(\frac{\partial \xi_k}{\partial x_j}\frac{\partial u_i}{\partial \xi_k}+\frac{\partial \xi_k}{\partial x_i}\frac{\partial u_j}{\partial \xi_k} -\frac{2}{3}\frac{\partial \xi_l}{\partial x_k}\frac{\partial u_k}{\partial \xi_l}\delta_{ij}  \bigg)
  \label{eq:diviatoric_stress}
\end{equation}
\begin{equation}
  \Theta_{i}=-\frac{1}{(\gamma - 1)M^2_\infty}\bigg(\frac{\mu}{Pr}\bigg)\frac{\partial \xi_j}{\partial x_i}\frac{\partial T}{\partial \xi_j}
  \label{eq:heat_flux}
\end{equation}
The $\hat{G}$ and $\hat{G_{\nu}}$ vectors have the same structure, with $\xi$ variations replaced by $\eta$ variations.
Stokes' hypothesis is assumed for the bulk viscosity coefficient~\textit{i.e} $\lambda=-2/3\mu$. 
Further details of the derivations may be found in~\citet{vinokur1989analysis}. 
Flow velocities and density are normalized by the free stream velocity~($u_\infty$) and density~($\rho_\infty$), respectively, time is non-dimensionalized by $c/u_\infty$~(where $c$ is the chord length of the airfoil), while the pressure is non-dimensionalized by 
$\rho_j U_\infty^2$; 
the perfect gas relationship then reads  $p=\rho T/\gamma M_\infty^2$, where $M_\infty$ refers to the freestream Mach number.
All plots shown in the paper are in non-dimensional units.
A constant Prandtl number of $0.72$ is assumed along with a constant ratio for the specific heats~$(\gamma =1.4)$.
Sutherland's law is used to model the temperature dependence of viscosity.

The Roe scheme is used for the inviscid fluxes~\cite{roe1981approximate} with a third-order reconstruction using the MUSCL approach~\cite{morrisonflux}. 
Viscous terms are discretized with a second-order central differencing technique. 
The temporal integration is performed explicitly using a $4^{th}$ order Runge-Kutta scheme with a time step size of $\Delta t=0.0001 c/u_\infty$.   
The use of the above methodology along with validation studies for flow past airfoils and cylinders can be found in \citet{ranjan2020robust}.

\section{Overall Flowfield Features}
\label{sec:flow_feat}

\begin{figure}
  \begin{subfigmatrix}{2}
    \subfigure[Computational Domain]{\includegraphics[width=0.48\linewidth]{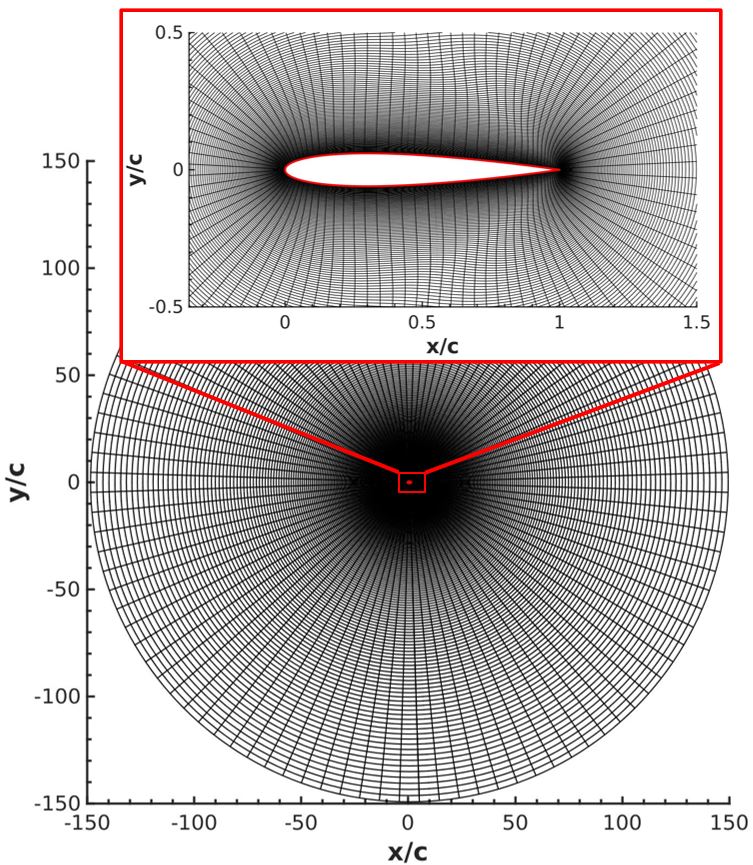}}
    \subfigure[Instantaneous snapshot]{\includegraphics[width=0.42\linewidth]{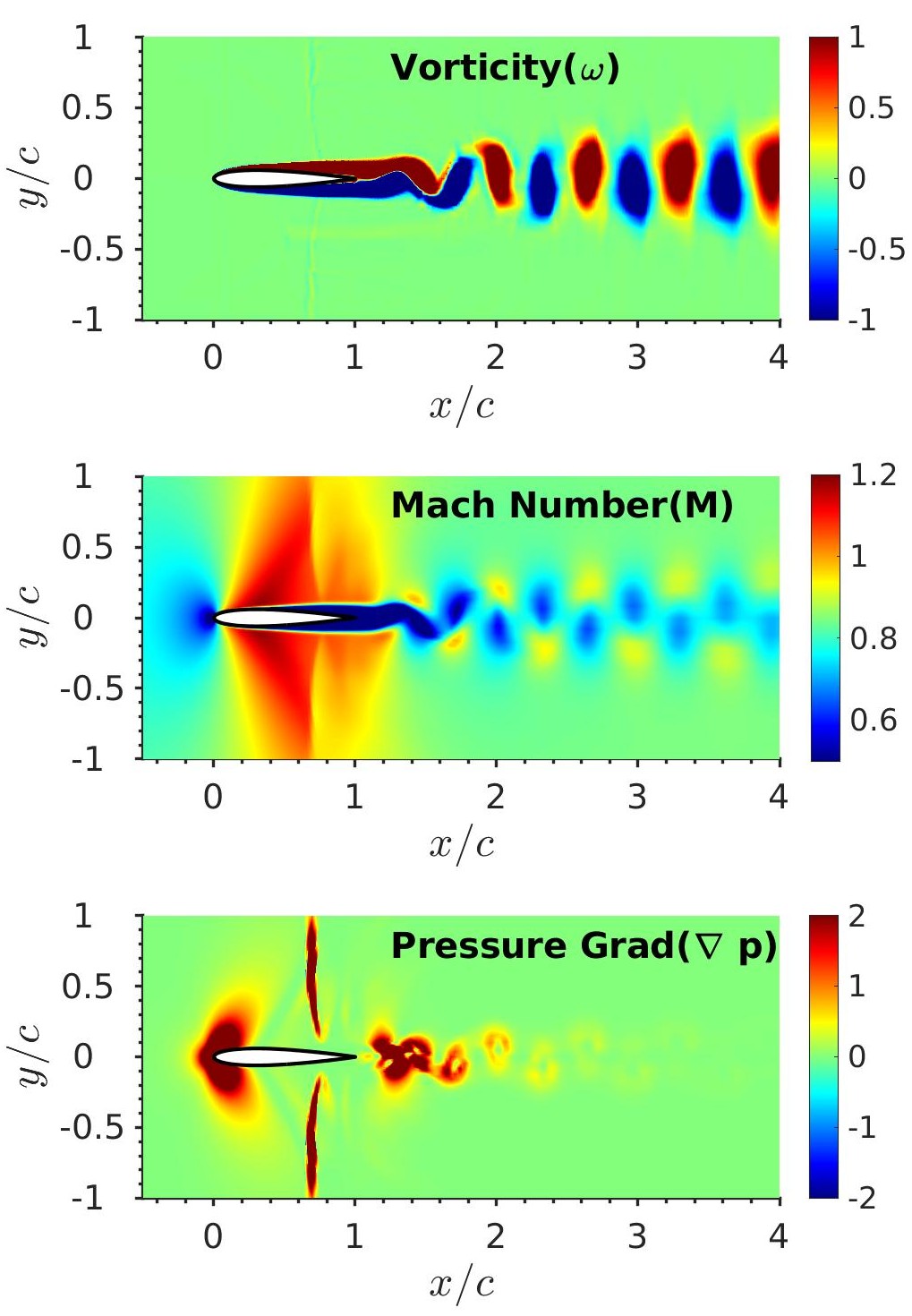}}
  \end{subfigmatrix}
  \caption{Details of the computational domain and flow physics shown through an instantaneous snapshot of the flowfield}\label{fig:overall_flow}  
\end{figure}

We choose a transonic flow past a NACA0012 airfoil at a zero angle of attack for the development of the ROMs. 
This flowfield has been widely explored using both numerical~\cite{bouhadji2003organised,bouhadji2003organised2} and experimental~\cite{harris1981two} techniques to analyze the unsteadiness in the wake including compressibility effects. 
At low Mach numbers~($M_\infty<0.7$), an undulation develops in the wake close to the trailing edge, which transitions to a pair of symmetrical recirculation bubbles with an increase in the freestream Mach number. 
In the transonic regime~($0.7<M_\infty<0.9$), a shock wave forms close to the trailing edge because of the acceleration of the flow up to supersonic speeds over the curved airfoil surface. The supersonic region is terminated by an unsteady shock wave, which induces a separation of the boundary layer and leads to the growth of a von K\'arm\'an instability. 
These instabilities result in an alternating vortex shedding resembling a von K\'arm\'an vortex street downstream of a cylinder.  


Following~\citet{bourguet2011reduced}, we choose a freestream Mach number of $0.85$ and a Reynolds number of $Re=5{,}000$ for the present simulation.
The transonic nature of the flow produces shock-induced unsteadiness in the airfoil wake. 
A stretched O-grid comprising $150{,}000$ points is used to discretize the domain, shown in fig.~\ref{fig:overall_flow}(a). 
The far-field boundaries are placed at $150$ chord lengths in all directions and are modeled using characteristic boundary conditions~\cite{whitfield1984three}. Figure~\ref{fig:overall_flow}b shows an instantaneous snapshot of the flowfield. 
The Mach number contours highlight the acceleration of the flow over the airfoil surface, giving rise to a region of supersonic flow around the airfoil. 
A shock forms towards the trailing edge, as seen in the $\nabla p$ contours, terminates this supersonic region. 
Vorticity contours display the growth of the von K\'arm\'an instability resulting in a vortex street in the wake. 
Shock oscillation and the coupled flapping shear layer induce strong pressure fluctuations resulting in significant entropic gradients in the flow. 
These gradients violate the assumptions for isentropic flow, which are only valid for low-to-moderate Mach numbers and have been instrumental in the derivation of several GP-based ROMs~\cite{rowley2004model}. 
This flowfield is thus a challenging testbed for evaluating the performance of the new NeuralGP algorithm, which has to date only been demonstrated for an isentropic flow over an open cavity~\cite{mohan2021learning}.






\begin{figure}
  \begin{subfigmatrix}{2}
    \subfigure[Normal turbulent stresses]{\includegraphics[width=0.6\linewidth]{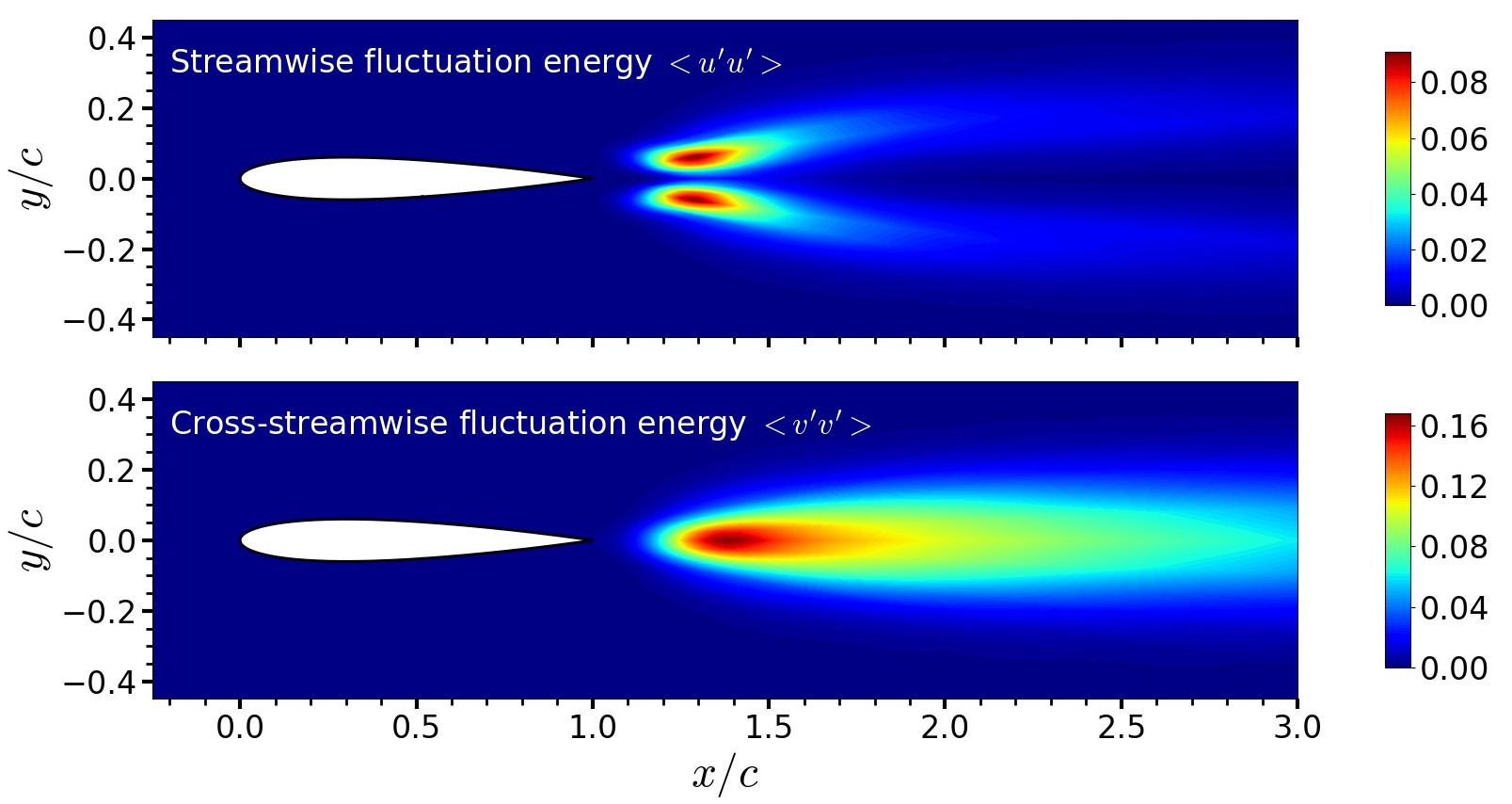}}
    \subfigure[Cross-streamwise velocity fluctuation ($v'$) spectra at $x/c=2.0$]{\includegraphics[width=0.39\linewidth]{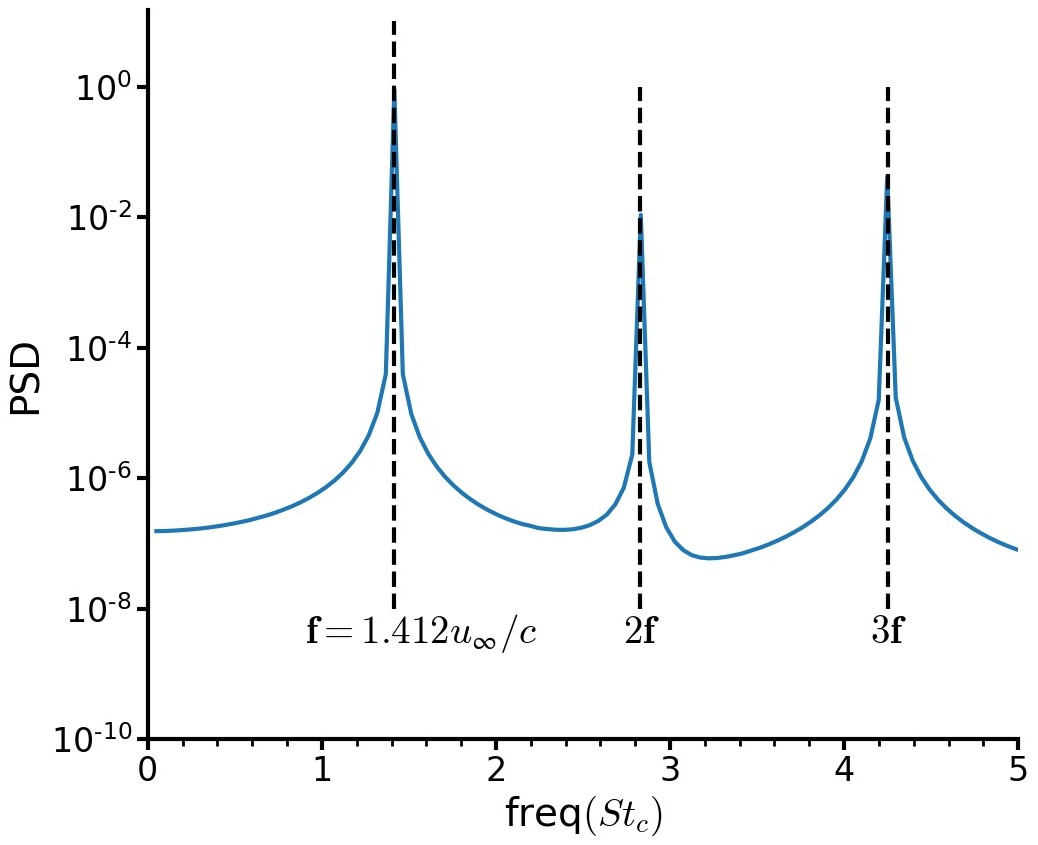}}
  \end{subfigmatrix}
  \caption{(a) Second order statistics of the flowfield shown using the normal turbulent stresses; (b) Power Spectral Density~(PSD) of velocity fluctuation spectra on the airfoil axis}\label{fig:SecondOrderStats}  
\end{figure}

The strength, spatial extent and period of the vortex shedding downstream of the airfoil are displayed using second order statistics.
Figure~\ref{fig:SecondOrderStats}(a). shows normal turbulent stresses downstream of the airfoil. 
The streamwise fluctuation energy~($\overline{u'u'}$) exhibits two symmetric lobes and is more concentrated close to the trailing edge of the airfoil while the cross-streamwise fluctuations~($\overline{v'v'}$) are more extended in the streamwise direction. 
Significant anisotropy is evident in the turbulent stresses with the cross-streamwise fluctuation energy~($\overline{v'v'}$) being an order of magnitude larger than the streamwise fluctuations. 
The peak values of $0.0912$ and $0.167$ for the streamwise and cross-streamwise normal stresses, respectively, are within $0.01$ of the values obtained in the benchmark simulations of~\citet{bouhadji2003organised2} for the same flow conditions. 
The frequency of the vortex shedding is assessed using the power spectrum of the cross-streamwise velocity fluctuations. 
Figure~\ref{fig:SecondOrderStats}b. shows the $v'$ spectrum on the axis of the airfoil one chord length downstream of the trailing edge~(at $x/c=2.0$). 
The fundamental vortex shedding frequency~(annotated as $f=1.412u_{\infty}/c$) and two super-harmonics dominate the spectrum. 
The fundamental shedding frequency lies within one frequency resolution bin of the values reported by~\citet{bouhadji2003organised2} for the same operating conditions, lending confidence in the accuracy of these simulations in capturing the dominant flow dynamics.

\section{Reduced Order Model}
\label{sec:ROM_method}


The ROM construction approach relies on a set of reduced basis functions optimized to represent the flowfield. 
Typically, POD modes are selected as the optimal basis~\cite{bourguet2011reduced,rowley2004model,peherstorfer2016data} owing to their ability to capture the maximum variance of the flow fluctuations and their mutual orthogonality, which simplifies the Galerkin projection of the governing equations.
The process used to compute the POD basis is outlined, followed by a description of the ROM construction.

\subsection{Proper Orthogonal Decomposition Procedure}
\label{sec:POD_process}

The following transformation of the primitive variables is applied prior to POD mode calculation:
\begin{equation}
  \{u,v,w,p,\rho\} \longrightarrow  q=\{1/\rho,u,v,w,p\}
\end{equation}
This facilitates the Galerkin projection of the full Navier-Stokes equations onto a reduced POD basis to construct the ROM~\cite{bourguet2011reduced}. 
The corresponding spatial POD basis functions~($\Phi_i=\{\Phi_i^{1/\rho},\Phi_i^u, \Phi_i^v,\Phi_i^w,\Phi_i^p\}^T$) are then obtained using $N_t$ snapshots of the flowfield sampled at $\Delta t= 5\times10^{-3}$. 
The snapshot POD method of \citet{sirovich1987turbulence} is used to compute the POD spatial modes, the details of which, including the spatial and temporal inner products employed, are provided for completeness in Appendix~\ref{sec:podDetails}. 

\begin{figure}
  \centering
  \includegraphics[width=0.75\linewidth]{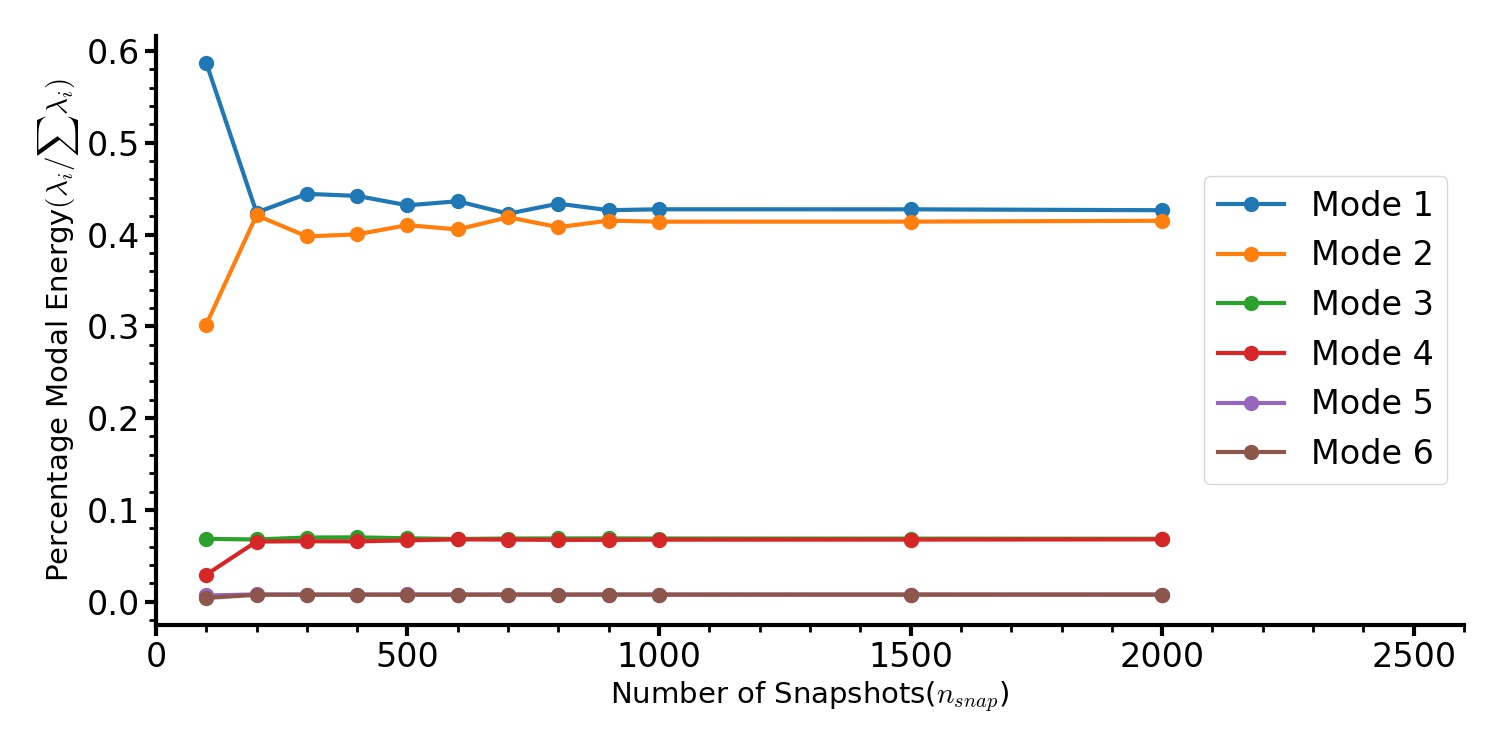}
  \caption{Effect of the number of snapshots used for the POD procedure on the modal energy for the leading six POD modes}
  \label{fig:POD_converge}
\end{figure}

The convergence of the POD modes is first established because of its substantial influence on the stability of the resulting ROMs.
This is achieved by examining the effect of the number of snapshots $N_t$ used for the POD procedure on the modal energies for the six leading POD modes, as shown in fig~\ref{fig:POD_converge}. 
For smaller values of $N_t$, the modal energies of the two leading POD modes are a strong function of the number of snapshots considered. 
This follows from the fact that the leading modes correspond to the lowest fundamental vortex shedding frequency~(fig~\ref{fig:SecondOrderStats}) and thus, fully converging their statistics requires snapshots distributed over a sufficiently long time horizon. 
With an increase in $N_t$, however, the modal energies of all six leading POD modes become invariant with respect to a change in $N_t$. 
Thus, to ensure an appropriately converged POD basis, we use $N_t=2{,}000$ snapshots to compute the POD modes.

\begin{figure}
  \centering
  \includegraphics[width=0.75\linewidth]{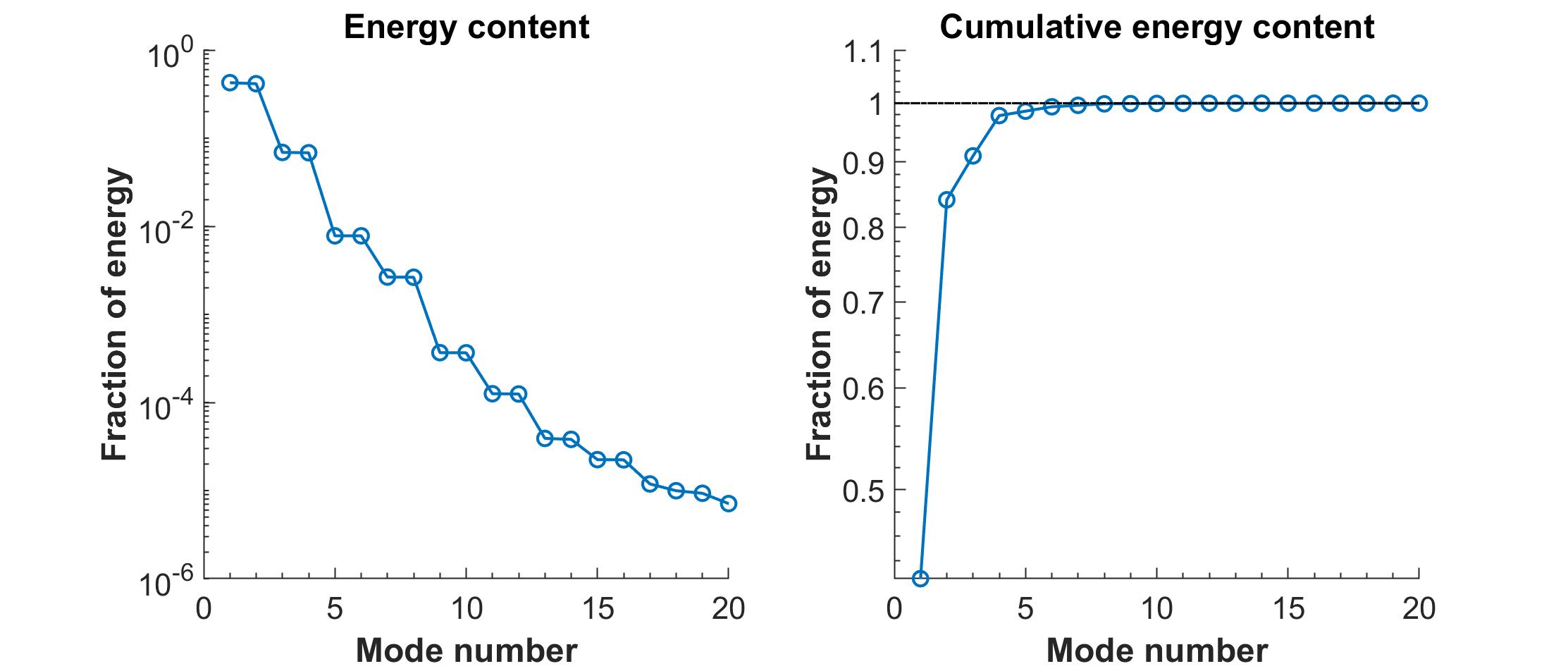}
  \caption{Fractional as well as the cumulative energy~($\sum_i^{N_{POD}} \lambda_i/\sum_i^{N_t} \lambda_i$) content of the individual POD modes given by eigenvalues of the cross correlation matrix}
  \label{fig:POD_energy}
\end{figure}

\begin{figure}
  \begin{subfigmatrix}{2}
    \subfigure[Streamwise velocity]{\includegraphics[width=0.49\linewidth]{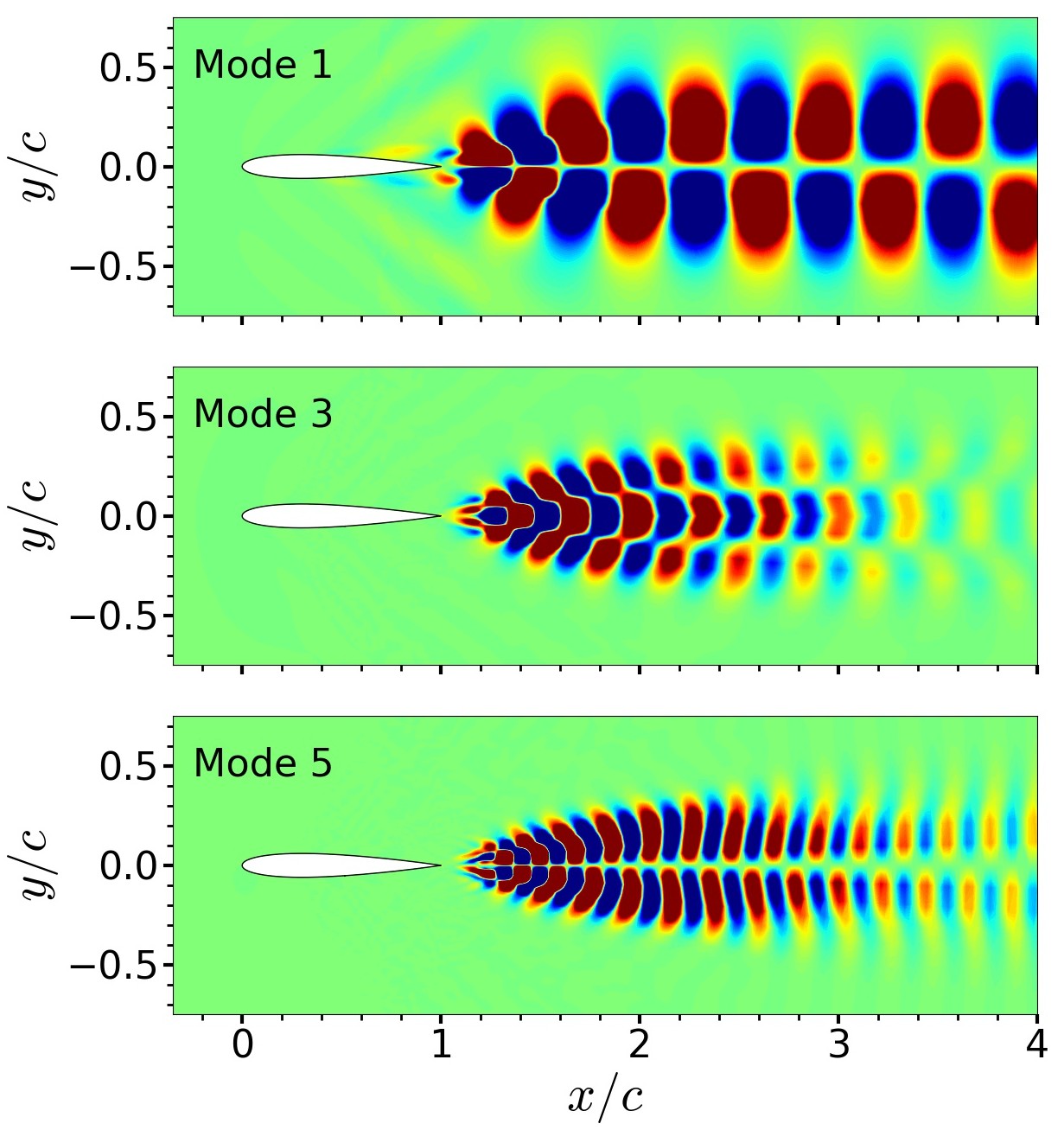}}
    \subfigure[Pressure]{\includegraphics[width=0.49\linewidth]{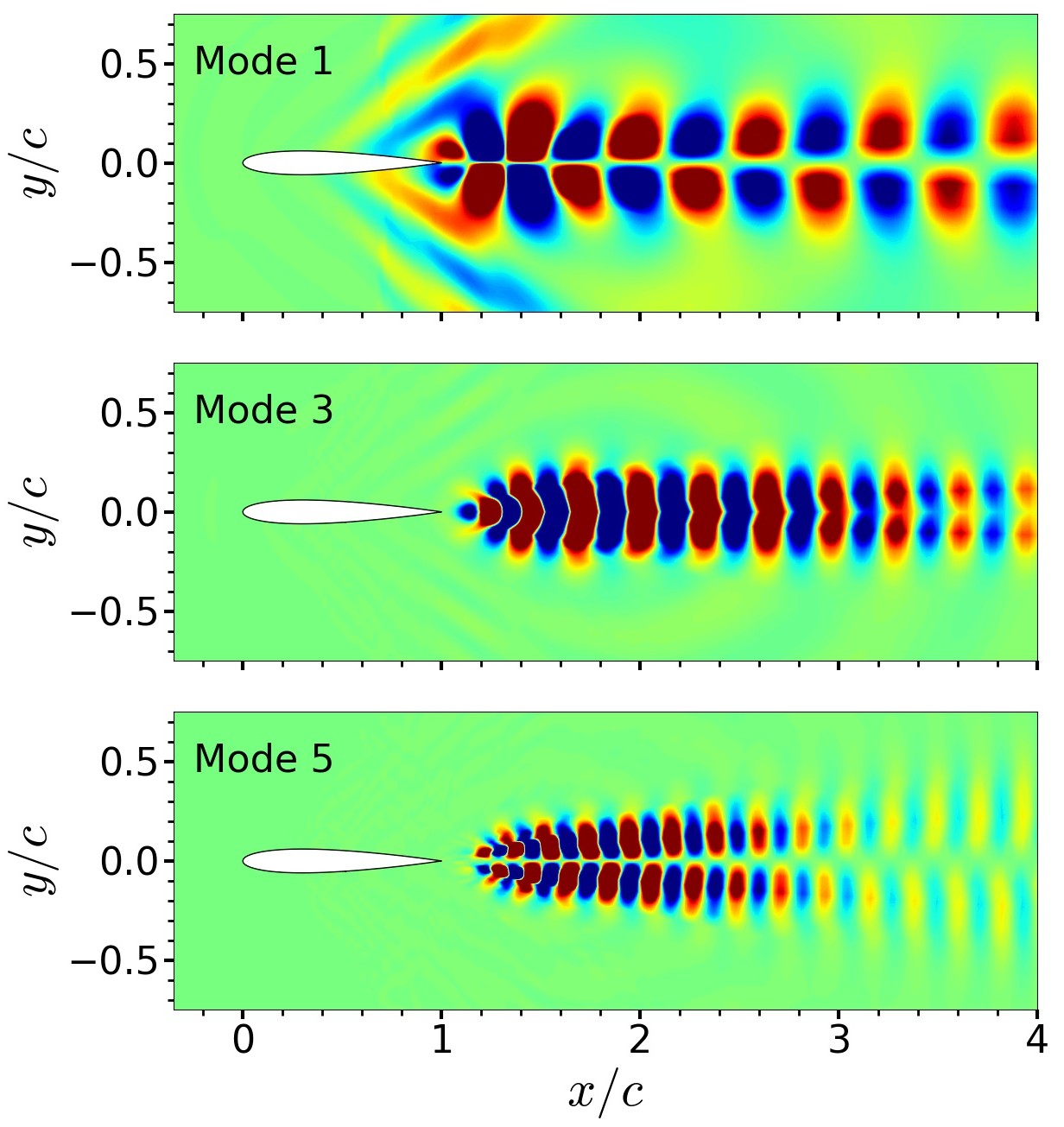}}
  \end{subfigmatrix}
  \caption{Leading spatial POD mode shapes for the 5 leading modes}\label{fig:POD_modes}  
\end{figure}

Figure~\ref{fig:POD_energy} shows the eigenvalues of the $20$ leading POD modes as a fraction of the energy in each mode, along with their respective cumulative energy content.
A rapid drop in the energy content of the higher POD modes is readily observable; as $99.3\%$ of the fluctuation energy is retained with a truncation beyond the 6 leading POD modes. 
Additionally, the leading modes occur in pairs, as is common in flows with dominant convective instabilities. 
Figure~\ref{fig:POD_modes}(a) and~(b) display the leading spatial POD modes for streamwise velocity and pressure fluctuations; for brevity, only odd numbered modes are shown among the repeated mode pairs.
The mode shapes represent convecting vortices in the wake, with the primary mode reflecting the spacing of the vortices in the von K\'arm\'an vortex street formed in the wake. 
Their antisymmetric pattern has been observed in previous studies on flows exhibiting a dominant von K\'arm\'an instability~\cite{bergmann2004optimal}. 
The higher modes represent increasingly finer spatial scales and correspond to superharmonics of the primary vortex shedding frequency. 

\subsection{Galerkin projection and calibration}
A Galerkin projection is performed of the full 3D compressible Navier-Stokes equations on the truncated POD basis comprising the leading $N_{POD}=6$ modes. 
The spatial inner product used for the projection of the equations is identical to that used for the classical POD algorithm~(detailed discussion in the context of equation~\ref{eq:in_prod_def} in Appendix~\ref{sec:POD_process}):
\begin{equation}
  \left< \frac{\partial\tilde{q}}{\partial t}, \Phi_i \right>_\Omega + \left< \frac{\partial\tilde{F}_j}{\partial \xi_j}, \Phi_i \right>_\Omega = \left< \frac{\partial\tilde{F}^{vis}_j}{\partial \xi_j}, \Phi_i \right>_\Omega 
\end{equation}
where the Einstein summation convention has been used. 
Galerkin projection yields the following set of ordinary differential equations for each time coefficient of the leading POD modes:
\begin{align}
  \begin{split}
    &\dot{a}_i = C_i + \sum_{j=1}^{N_{pod}} L_{ij}a_j + \sum_{j=1}^{N_{pod}} \sum_{k=1}^{N_{pod}} Q_{ijk} a_j a_k \\ 
    &a_i(0) = \left<\tilde{q}_0-\overline{\tilde{q}}, \Phi_i\right>_\Omega
  \end{split}
  \label{eq:ROM_ODE}
\end{align}

As discussed previously, the time integration of the above model suffers from instabilities~\cite{iollo2000stability,balajewicz2012stabilization,amsallem2012stabilization,akhtar2009stability} resulting in a significant prediction error, alteration of oscillation periods, and, finally, a divergence of the temporal integration. 
To obtain a stable prediction that minimizes prediction errors, in the present study we implement a calibration method similar to that used by~\citet{bourguet2011reduced}. 
The calibration procedure modifies the behavior of the ROM equations using calibration matrices ($L_{ij}^{cal}$ and $C_{i}^{cal}$) added to the original ROM coefficient matrices~($L_{ij}$ and $C_{i}$).
This results in the following equation:
\begin{equation}
  \dot{a}_i = \left( C_i + C_{i}^{cal} \right) + \sum_{j=1}^{N_{pod}} (L_{ij}+L_{ij}^{cal})a_j + \sum_{j=1}^{N_{pod}} \sum_{k=1}^{N_{pod}} Q_{ijk} a_j a_k = f_i(L^{cal},C^{cal},\mathbf{a})
  \label{eq:ROM_ODE_cal}
\end{equation}
We obtain the calibration matrices by solving an optimization problem that seeks to minimize the ROM prediction error subject to the constraint of calibration cost. 
Thus, the following cost function is minimized:
\begin{equation}
  \mathcal{J}(C^{cal},L^{cal},\theta) = \theta\mathcal{E}(C^{cal},L^{cal}) + (1-\theta)\mathcal{C}(C^{cal},L^{cal})
  \label{eq:J_def}
\end{equation}
where $\mathcal{E}(C^{cal},L^{cal})$ and $\mathcal{C}(C^{cal},L^{cal})$ represent normalized ROM error and calibration cost, respectively. 
The two terms are blended using the coefficient $\theta \in \left( 0,1 \right)$. 

The choice of error metric and calibration cost has a significant bearing on the calibrated ROM equations and merit a discussion. The ROM prediction error is defined as:
\begin{equation}
  E(C^{cal},L^{cal}) = \sum_{i=1}^{N_{pod}}\int_0^T \left( a_i^{pod}(t)-a_i^{pod}(0)-\int_0^t f_i(C^{cal},L^{cal},a^{pod})dt' \right)^2 dt.
  \label{eq:err_def}
\end{equation}
Here, $a_i^{pod}$ indicates the time coefficient of the $i^{th}$ POD mode observed directly from the flow field, referred to hereafter as the reference POD time coefficient. 
Thus, the ROM prediction error is calculated as the difference between the change in the $i^{th}$ reference POD time coefficients~($a^{POD}_i$) and that predicted by the RHS of the ROM equation when using the reference POD time coefficients~\ref{eq:ROM_ODE_cal} over a time horizon $t=0$ to $t$.
Alternative definitions for the ROM prediction error, such as the difference between the instantaneous time rate of changes of $a_i$s and that predicted by the RHS of the governing equation, may also be used~\cite{cordier2010calibration}:
\begin{equation}
  \label{eq:1}
  E^{alt}(C^{cal},L^{cal})=\sum_{i=1}^{N_{pod}} \int_0^T \left(\left(\frac{d a_i^{pod}}{dt} - f(C^{cal},L^{cal},a^{pod})\right)^{2}dt\right)
\end{equation}
Crucially, these expressions represent a linearized definition of the ROM error which only consider the differences in the rate of change of POD coefficients~($\dot{a}$) versus that predicted by the ROM \textit{at each time step}. 
While this choice results in a linearization of the optimization problem, it ignores the cumulative effects of the small differences in the predicted $\dot{a}$ over the entire calculated trajectory. 
The consequences of this choice are analyzed in greater detail in section~\ref{sec:sysDynAnalysis}.  

The error term~$E$ is normalized using the ROM prediction error with no calibration to obtain $\mathcal{E}$:
\begin{equation}
  \mathcal{E}(C^{cal},L^{cal}) = \frac{E(C^{cal},L^{cal})}{E(0,0)}
  \label{eq:err_norm}
\end{equation}
The calibration cost is defined as a measure of the calibration matrices relative to the system dynamic matrices. 
Here, a sum of an $L_2$ norm of the $C_i$ vector and a Frobenius norm of the $L_{ij}$ matrix is considered:
\begin{equation}
  \mathcal{C}(C^{cal},L^{cal}) = \frac{||C^{cal}||^2+||L^{cal}||^2}{||C||^2+||L||^2}
  \label{eq:calib_def}
\end{equation}
The calibration procedure then solves the optimization problem to minimize $\mathcal{J}$ for a given value of $\theta$ resulting in specific calibration matrices $L^{cal}$ and $C^{cal}$ obtained as a function of the blending parameter $\theta$.

Minimizing $\mathcal{J}$ is equivalent to solving the following linear system of size $N_{POD}+1$:
\begin{equation}
  A^{cal}(K^{cal}_i)^T=b^i
\end{equation}
where $K^{cal}=[C^{cal},L^{cal}]$ and the matrices $A^{cal}$ and $b^i$ are given as:
\begin{eqnarray}
  A_{ij}&=&\int_0^T \left( \int_0^t a_i^\star dt' \right) \left( \int_0^t a_j^\star dt' \right) dt + \theta^\star\delta_{ij} \\
  b_i^k &=& \int_0^T \left( a_k^{pod}(t)-a_k^{pod}(0)-\int_0^t f_k \left(0,0,a^{pod} \right) dt'	\right) \left( \int_0^t	a_i^\star dt' \right) dt,
  \label{eq:Ab_def}
\end{eqnarray}
with $a^\star=[1,a_1,a_2,\cdots,a_{NPOD}]$ and 
\begin{equation}
  \theta^\star=\frac{1-\theta}{\theta} \frac{E(0,0)}{||C||^2+||L||^2}
\end{equation}


\subsection{Neural Galerkin ROM using Differentiable Programming}
The GP formulation of equation~\ref{eq:ROM_ODE} gives rise to $C$ (Constant), $L$ (Linear) and $Q$ (Quadratic) terms, that are then modified for greater stability via an optimization and calibration process \textit{from the data}. 
We remark that the data-driven nature of eq.~\ref{eq:ROM_ODE_cal}, while being structured as an ODE, makes it a good candidate to be cast instead as a DiffProg problem. 
Specifically, the $C$ and $L$ coefficients have a significant impact on the accuracy and long-time stability of the GP-ROM~\citep{nagarajan2013development,bourguet2011reduced}.
Since this entails considerable manual intervention as calibration, we seek to automate the discovery of these coefficients with DiffProg, and directly use the trajectories of the reduced state~($a_i(t)$) from the high resolution simulation as the training data. 

In the DiffProg problem setup, the ``unknown" operators in  equation~\ref{eq:ROM_ODE} are approximated with trainable parameters $p$ in a Neural Network~(NN), such that $C_i \approx C_i^{p}$ and $L_{ij}  \approx L_{ij}^{p}$, so that the equation is now,
\begin{align}
  \begin{split}
    &\dot{a}_i = C_{i}^{p} + \sum_{j=1}^{N_{pod}} L_{ij}^{p} a_j + \sum_{j=1}^{N_{pod}} \sum_{k=1}^{N_{pod}} Q_{ijk} a_j a_k \\  
   &a_i(0) = \langle\tilde{q}_0-\overline{\tilde{q}}, \Phi_i\rangle_\Omega
  \end{split}
  \label{eq:neuralGP_ODE}
\end{align}
Since the trainable parameters $p$ have to be learned, they are typically initialized with random numbers, as is standard in deep learning literature~\cite{narkhede2022review}. 
In the forward pass of our setup, we solve Eqn.~\ref{eq:neuralGP_ODE} using standard ODE solvers to estimate $a_i(t)$ trajectories over a stipulated time period~($T$). 

The first estimate of the predicted trajectories will necessarily be inaccurate since the operators are essentially random, and the rest of the ODE propagates it in time. 
This error in trajectory is captured in the loss function, which is defined as the $L_2$ norm of the error between the calculated trajectories and those obtained from the high-resolution simulation:  
\begin{align}
  L(\mathbf{a}(t_0))&= \sum_{i=1}^{N_{POD}}\int_0^{T} \bigg\{{a_i^{ref}}(t_1)-\int_{0}^{t_1} f_i(\mathbf{a},t)dt\bigg\}^2 dt_1    \label{eq:lossDef}\\
  &=MSE(ODESolve(\mathbf{a}(t_0),t_0,f,\theta)-\mathbf{a}_{ref}) 
\end{align}
where $f_i(\mathbf{a}, t)$ represents the RHS of Eqn.~\ref{eq:neuralGP_ODE} and $a_i^{ref} $ refers to the reference POD time coefficients. 
Crucially, the solution of the ODE problem with given parameters $C_{i}^{p}$ and $L_{ij}^{p}$ is an integral part of the loss function and the prediction error is calculated for the entire trajectory over a time period~(from $t=0$ to $T$). 
In the backward pass, the NN parameters $p$ are updated by backpropagating the error through the ODE. 
The updated $p$ is slightly more accurate than the previous iteration and used in the forward pass for the next iteration, where the loss is computed again. 
This forward-backward pass cycle is repeated till the error has reduced to a satisfactory value, which is user-defined and based on the accuracy requirements of the application. 
We outline this process in Fig.~\ref{fig:NeuralGPgraphic}, as it has some distinct advantages compared to other methods in literature:
\begin{figure}
  \centering
  \includegraphics[width=\linewidth]{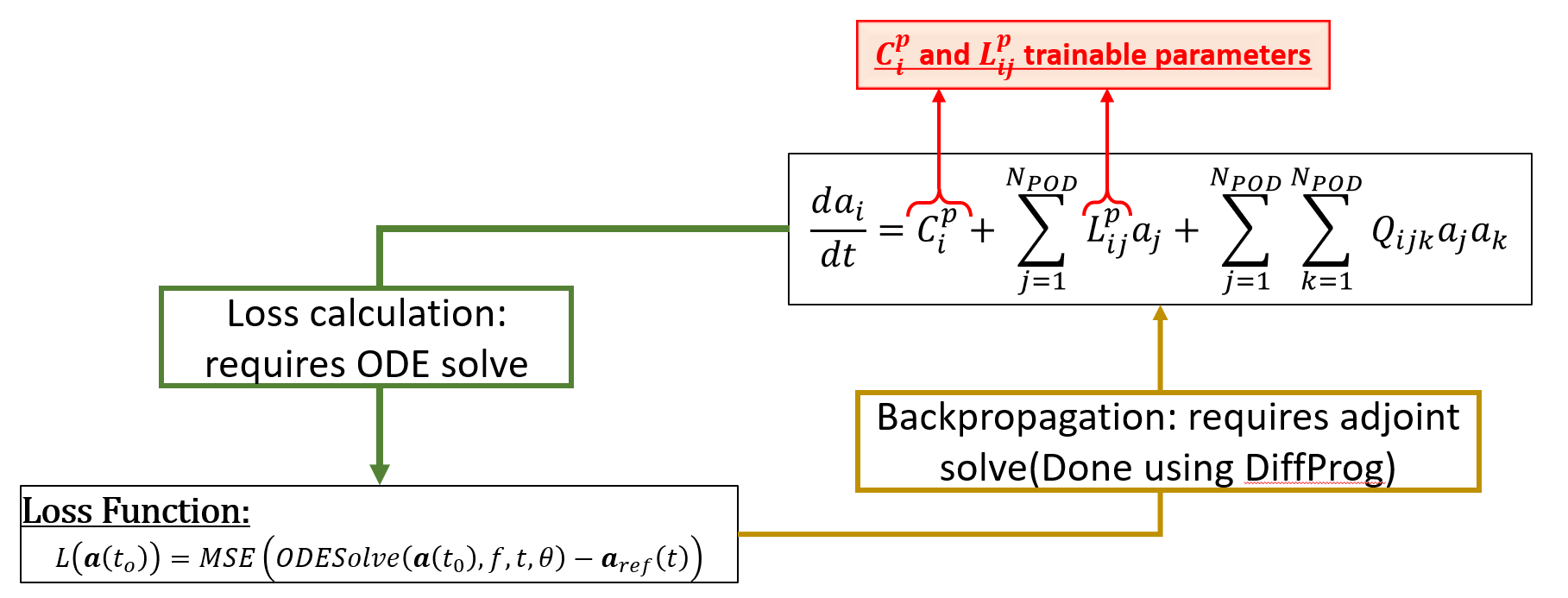}
  \caption{Schematic of NeuralGP training by treating the linear and constant ROM-ODE coefficients as trainable parameters}
  \label{fig:NeuralGPgraphic}
\end{figure} 

\begin{itemize}
\item Both $C_{i}$ and $L_{ij}$ are treated as trainable parameters in the NN and they interact in a \textit{physical manner} governed by the form of the projected Navier-Stokes equations.

\item DiffProg allows for NN models to be back propagated seamlessly thru the ODE solver via  adjoint sensitivity analysis and a combination of forward/reverse mode Automatic Differentiation (AD). 
The resulting abstract, powerful framework allows for arbitrary neural network models to be tightly integrated within the context of the well-known GP equations.
\end{itemize}
Thus, contrary to the previously adopted deep learning approaches for for fluid flows~\cite{mohan2018deep,shen2021dynamic,deng2019time,srinivasan2019predictions} that use black-box neural networks employing standardized non-linearities, the NeuralGP approach leverages the power of the DiffProg paradigm to derive coefficients to a known non-linearity~(the Galerkin ROM-ODE).

Our method may thus be viewed as an extension to the non-intrusive ROM techniques proposed by \citet{peherstorfer2016data,ghattas2021learning} which seek to obtain the physically consistent operator matrices for a known form of the ROM.
These methods use training data composed of ROM trajectories calculated directly from the full-order model sampled at discrete points and solve a linear optimization problem to obtain the ROM operator matrices. 
In contrast, in the current approach, the operators are learned by minimizing a more rigorous prediction error defined over the entire trajectory obtained by integrating the ROM. 
This rigorous definition of the error results in robust learned data-driven models while retaining the key advantages of physically derived ROM ODEs~\cite{ghattas2021learning}. 
Two features are noteworthy.  
First, the computational cost of training is very low since retention of physics as its Galerkin ODE structure allows leveraging of fast, cheap ODE solvers instead of expensive, complex deep NN architectures. 
These deep networks,  like LSTMs and CNNs, are black boxes that need to represent the entire physics, which requires significant computing resources (several GPUs) to learn even modest-sized datasets~\citep{mohan2018deep,mohan2019compressed,mohan2020spatio}. 
Second, the algorithm attempts to learn a physically consistent system governing equation rather than fitting the data to a generic model for prediction; this alleviates concerns regarding over-training the neural network for a dataset at the cost of generalizability. 
As noted earlier, the use of differentiable programming approach has demonstrated substantial advantages for isentropic flows over Galerkin projection/calibration methods~\cite{mohan2021learning}.   
Here we demonstrate that the method shows similar improvement for more general flows where isentropic behavior is not observed.

\section{Evaluation of NeuralGP-based ROM}
\label{sec:results}

The performance of the proposed approach in obtaining stable ROM coefficients is evaluated using short term and long-term predictions. 
The method uses reference POD coefficients over a time horizon of $t=1.5 c/u_\infty$, corresponding to two full periods of the leading POD mode.
For testing, the calibrated ROM was also obtained using the same time horizon and a blending parameter of $\theta=0.9$, corresponding to a $2.25\%$ calibration cost~(eq.~\ref{eq:calib_def}). 
The GP-ROM equations are shown in appendix~\ref{sec:GPROMDetails}.
The NeuralGP method underwent $2{,}000$ training iterations to reduce the loss function by five orders of magnitude, which required $\approx 228$ seconds on an Intel i7-8550U octa-core CPU running at $4$ GHz.
The inference cost for both the calibrated GP-ROM as well as the NeuralGP ROM are similar, since they both require an ODE-solve.


\subsection{Short time prediction}

\begin{figure}
  \centering
  \includegraphics[width=\linewidth]{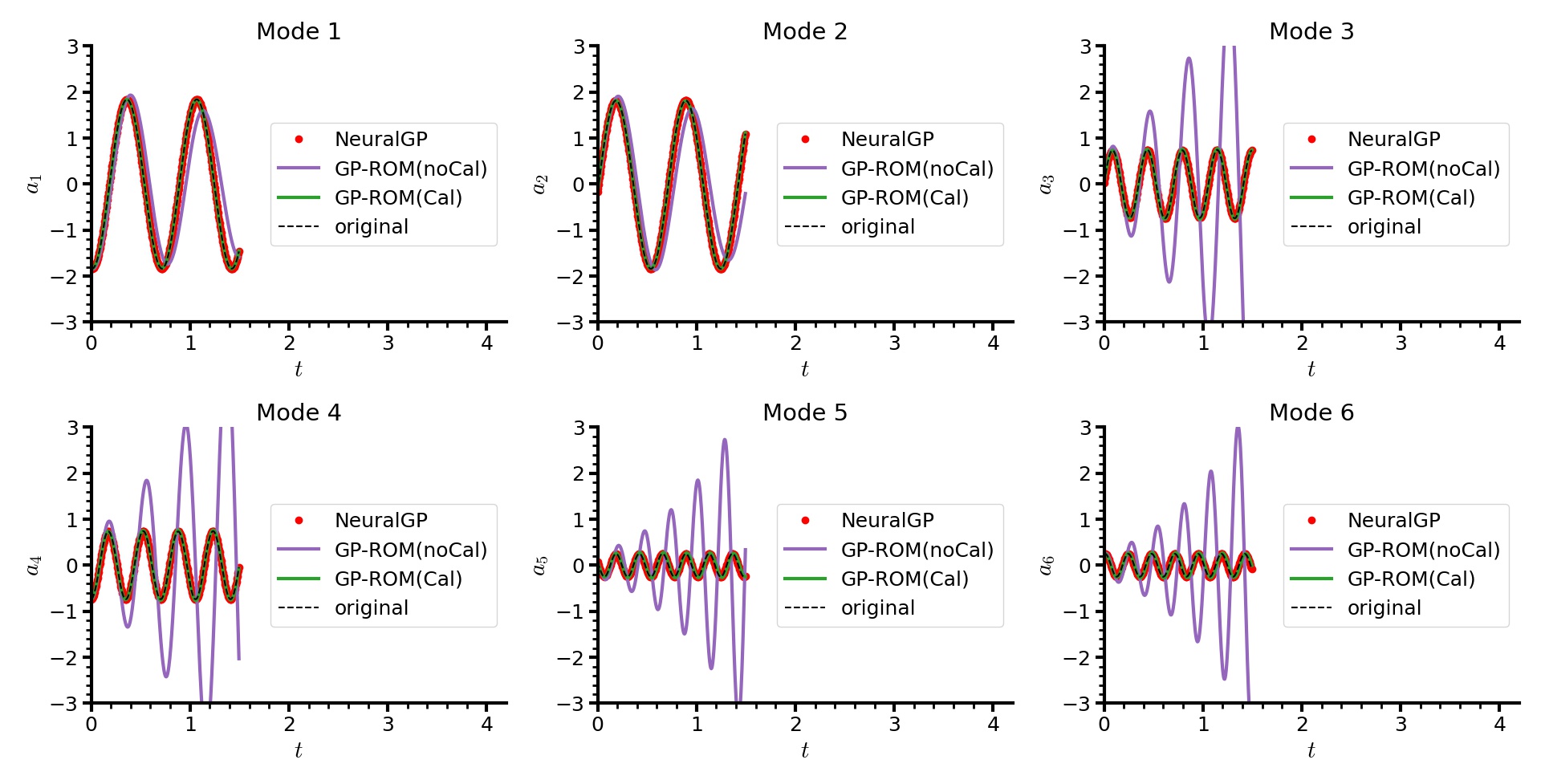}
  \caption{Reference time coefficients~$a_i(t)$ of the 6 leading POD modes (original) along with predictions from a) calibrated ROM (Cal), b) Uncalibrated ROM (noCal) and c) NeuralGP ROM, obtained using Galerkin Projection. Uncalibrated ROM quickly diverges compared to the other two approaches.}
  \label{fig:ROM_results}
\end{figure}


Figure~\ref{fig:ROM_results} shows the time coefficients~$a_i(t)$ for the six leading POD modes, along with predictions from the uncalibrated GP-ROM, calibrated GP-ROM, and the NeuralGP-ROM. 
Time coefficients of all leading modes exhibit a sinusoidal variation with the repeated pairs of eigenvalues discussed in fig.~\ref{fig:POD_energy} having the same period but with a $\pi/2$ phase lag to maintain their orthogonality in time. The frequencies corresponding to the time coefficients also increase with the mode number, along with a progressive reduction in amplitude. 

The uncalibrated ROM diverges relatively quickly from the reference POD time coefficients, especially in the higher POD modes, with rapid growth in amplitudes and a change of the oscillation periods. 
Even for the two leading POD modes, reasonable results are only obtained over the first oscillation period, after which we can observe a significant deviation from the reference POD coefficients. 
Even with the relatively small additions to the $L_{ij}$ and $C_i$ matrices~(as indicated by the small calibration cost), the calibrated ROM tracks the time coefficients obtained from the simulation accurately over the short time horizon. Finally, the results of the NeuralGP ROM closely track the original POD time coefficients, resulting in virtually indistinguishable differences between the calibrated GP and NeuralGP ROM results from the original POD time coefficients in the short time horizon. 
While the close correspondence of the ROM predictions with the reference POD time coefficients in the short time horizon is encouraging, it only serves as a sanity check for the ROM methodologies. 
This is because the short time horizon over which the predictions are being tested here is identical to that used for calibrating and training the GP and NeuralGP ROMs, respectively. 
We now undertake a more rigorous test of the ROMs by evaluating their performance over a longer term prediction horizon.

\subsection{Long time predictions}

\begin{figure}
  \begin{minipage}{\linewidth}
    \centering
    \subfigure[Long term predictions 
    ]{\includegraphics[width=0.99\linewidth]{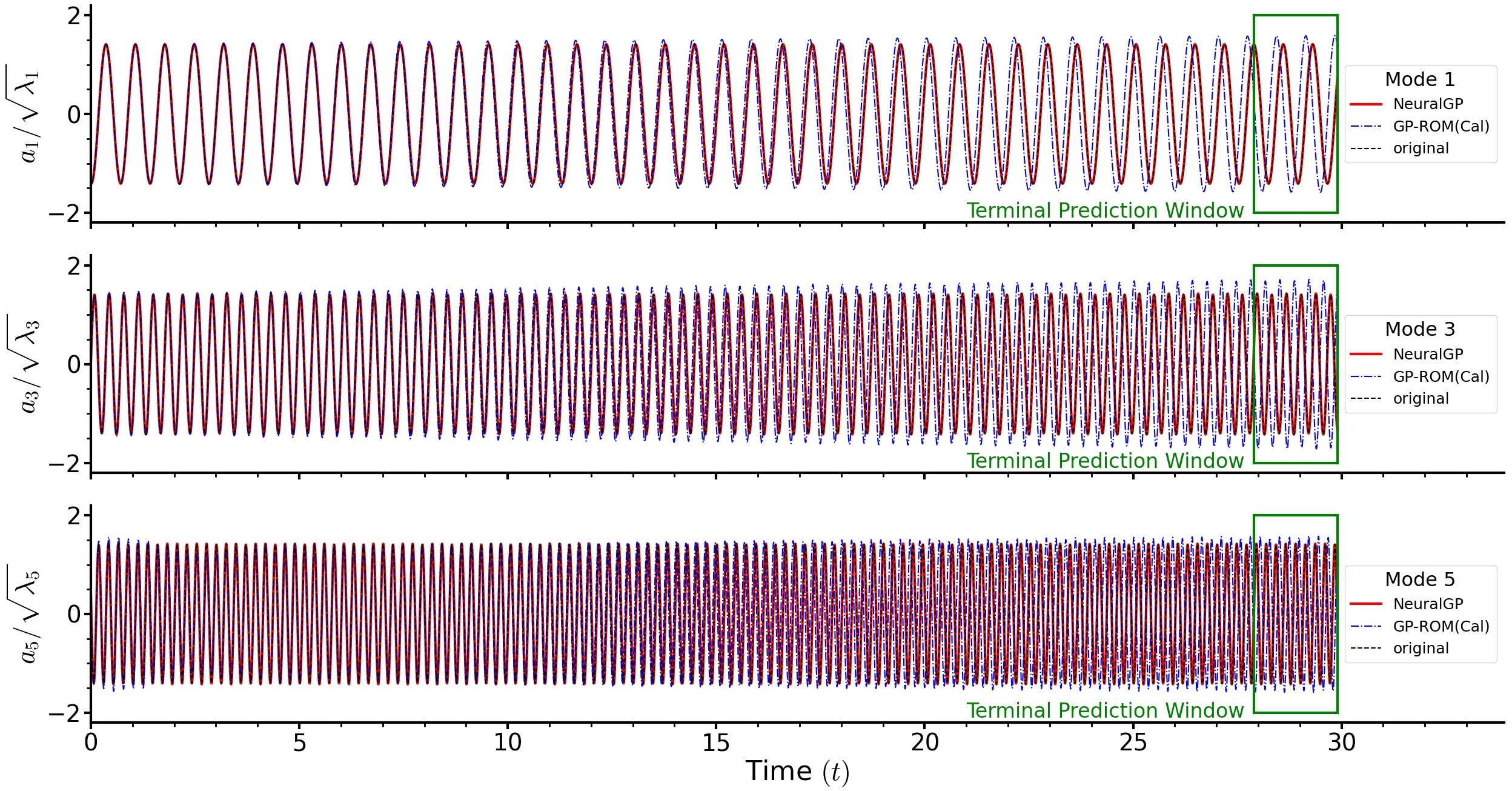}}
  \end{minipage}
  \begin{minipage}{.49\linewidth}
    \centering
    \subfigure[Zoomed in predictions in the terminal prediction window]{\includegraphics[width=\linewidth]{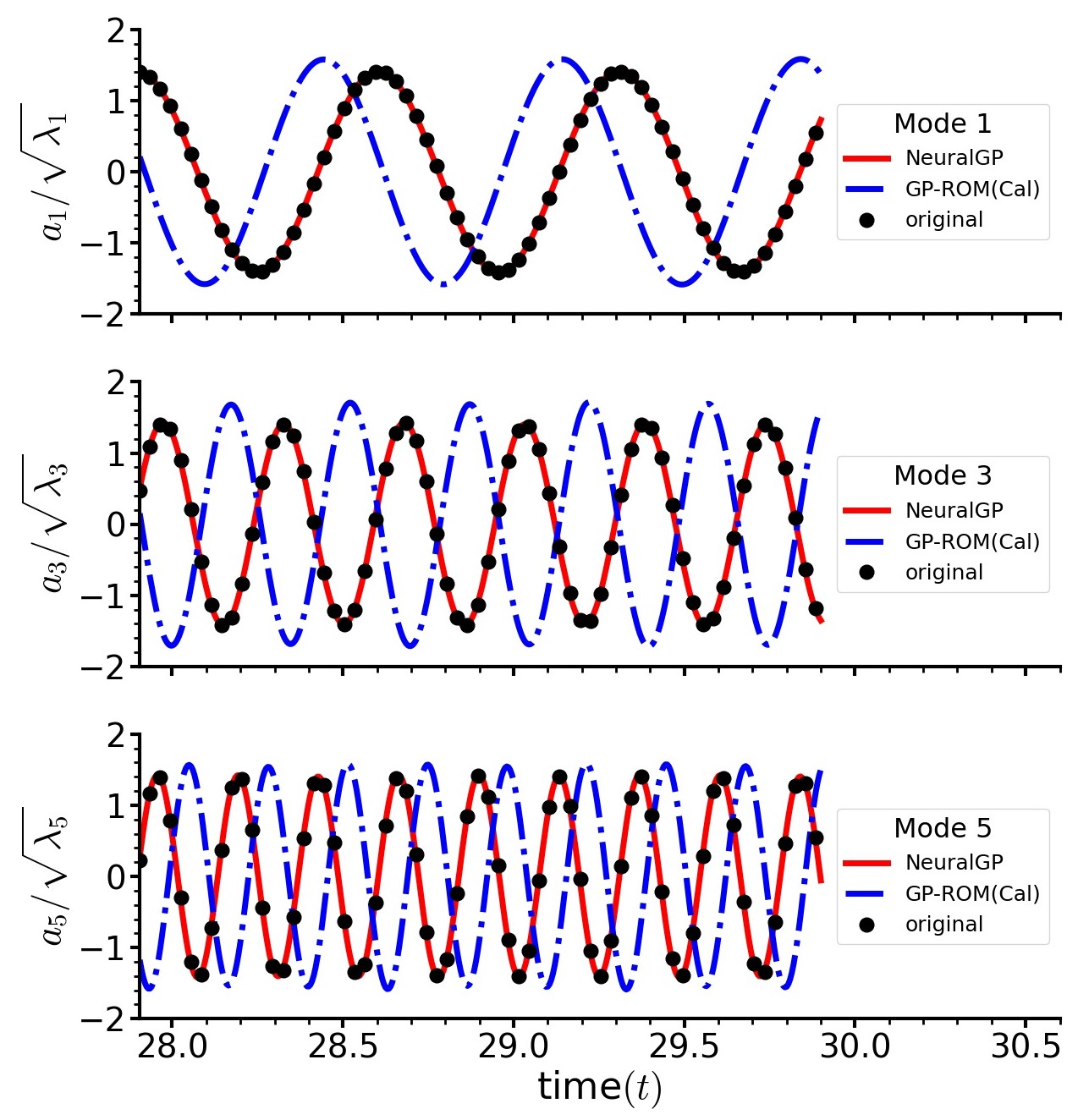}}
  \end{minipage}
  \begin{minipage}{.49\linewidth}
    \centering
    \subfigure[Spectral distribution at the beginning and end of the prediction window with the fundamental shedding frequency($St=1.42$) and its harmonics being identified in each pair of POD modes] {\includegraphics[width=\linewidth]{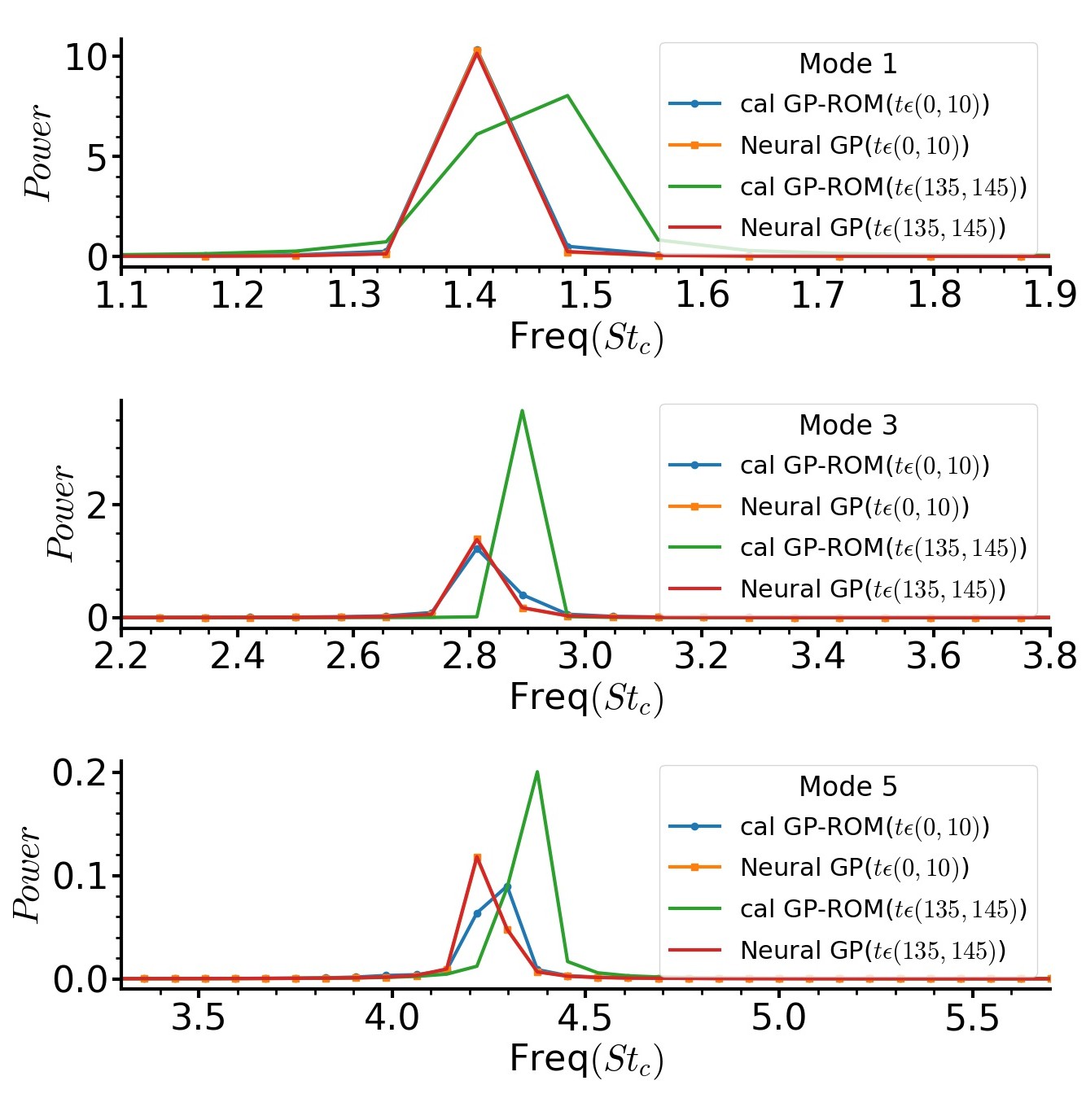}}
  \end{minipage}
  \caption{Long time prediction of the 4 leading POD modes using the calibrated GP-ROM and NeuralGP ROMs along with a detailed view of the terminal prediction window}
  \label{fig:longTimeForecasts}
\end{figure}

The long term predictive capabilities of both ROMs are evaluated by integrating the respective ROM-ODEs over a longer time horizon ($t \in (0,30 c/u_\infty)$).
Figure~\ref{fig:longTimeForecasts}(a) shows the energy-normalized long time predictions of the three leading odd number modes obtained using both the calibrated GP and NeuralGP ROMs compared to the original POD time coefficients.
The even modes exhibit similar behaviors as their corresponding odd mode pairs, and are therefore not displayed here for brevity.
The predictions from both ROMs follow the original POD time coefficients closely until $t\approx10 c/u_\infty$. 
Beyond this time, the NeuralGP predictions track the original POD time coefficients closely, with no indication of divergence, though the GP-ROM shows significant errors in prediction. 
The terminal prediction window at the end of the long time prediction horizon is shown in figure~\ref{fig:longTimeForecasts}b.
There is relatively little degradation in the performance of the NeuralGP while 
the calibrated GP-ROM chiefly changes in the oscillation periods, along with a slow growth in the amplitude. 
These gradual changes in the frequencies of each time coefficient lead to substantial errors in the predicted phase of the modes, as observed for modes $3$ and $5$ in the terminal prediction window.  

The performance of the NeuralGP ROM is further investigated using Power Spectral Densities~(PSD) of the time coefficients.
To understand the differences in the spectral content brought about by the longer term integration of the ROMs, the analysis is performed on smaller time windows of the long-time prediction horizon~($tU_\infty/c \in (0,150 )$).
Figure~\ref{fig:longTimeForecasts}(c) shows the differences in the spectral content of the three leading mode pairs within time windows of $10 c/U_{\infty}$ at the start ($tU_{\infty}/c\in (0,10)$) and end ($tU_{\infty}/c\in (135,145)$) of the long time prediction horizon.
The leading POD modes captures the fundamental shedding frequency marked in fig.~\ref{fig:SecondOrderStats}(b) while the superharmonics are captured by the progressively higher POD mode pairs.
The extents of the $x$-axes of the PSD plots are adjusted to represent the correspondingly higher temporal frequencies of the higher POD time coefficients. 
The GP-ROM prediction shows a substantial difference in the spectral content over the course of the entire prediction horizon.
The calibrated GP-ROM yields much better results, but there is nonetheless a shift in the frequency content and amplitude growth in the time window close to the end of the long time prediction horizon~$t \in (135,145)$.
The NeuralGP predictions maintain a near identical spectral distribution matching closely with the spectral signature of the vortex shedding~(fig.~\ref{fig:SecondOrderStats}(b)) irrespective of the time window selected. This appears to be a systematic characteristic of NeuralGP, since it minimizes the full trajectory error in a nonlinear fashion. We study this in detail in the next section.

\section{System dynamics analysis of the ROM-ODE}
\label{sec:sysDynAnalysis}

The attractive properties of the long term predictions of the NeuralGP-ROM are now traced to aspects of its underlying mathematical formulation.
As mentioned in the previous section, the calibration procedure alters the ROM coefficients obtained from the Galerkin projection of the Navier-Stokes equations on the POD basis 
to stabilize the ROM predictions.
This is achieved by solving an optimization problem seeking to minimize a cost function~($\mathcal{J}$ defined in eq.~\ref{eq:J_def}), which is composed of a linearized definition of the prediction error~($\mathcal{E}$ defined in eq.~\ref{eq:err_norm}) and a calibration cost penalty~($\mathcal{C}$ defined in eq.~\ref{eq:calib_def}), balanced via a blending parameter~($\theta$). 
These parameters dictate the properties of the final calibrated ROM dynamic matrices, therefore studying their variations can provide important insights into the calibration process itself. 


\begin{figure}
  \begin{subfigmatrix}{2}
    \subfigure[Full Range of $\theta$($\theta \in (0,1)$)]{\includegraphics[width=0.63\linewidth]{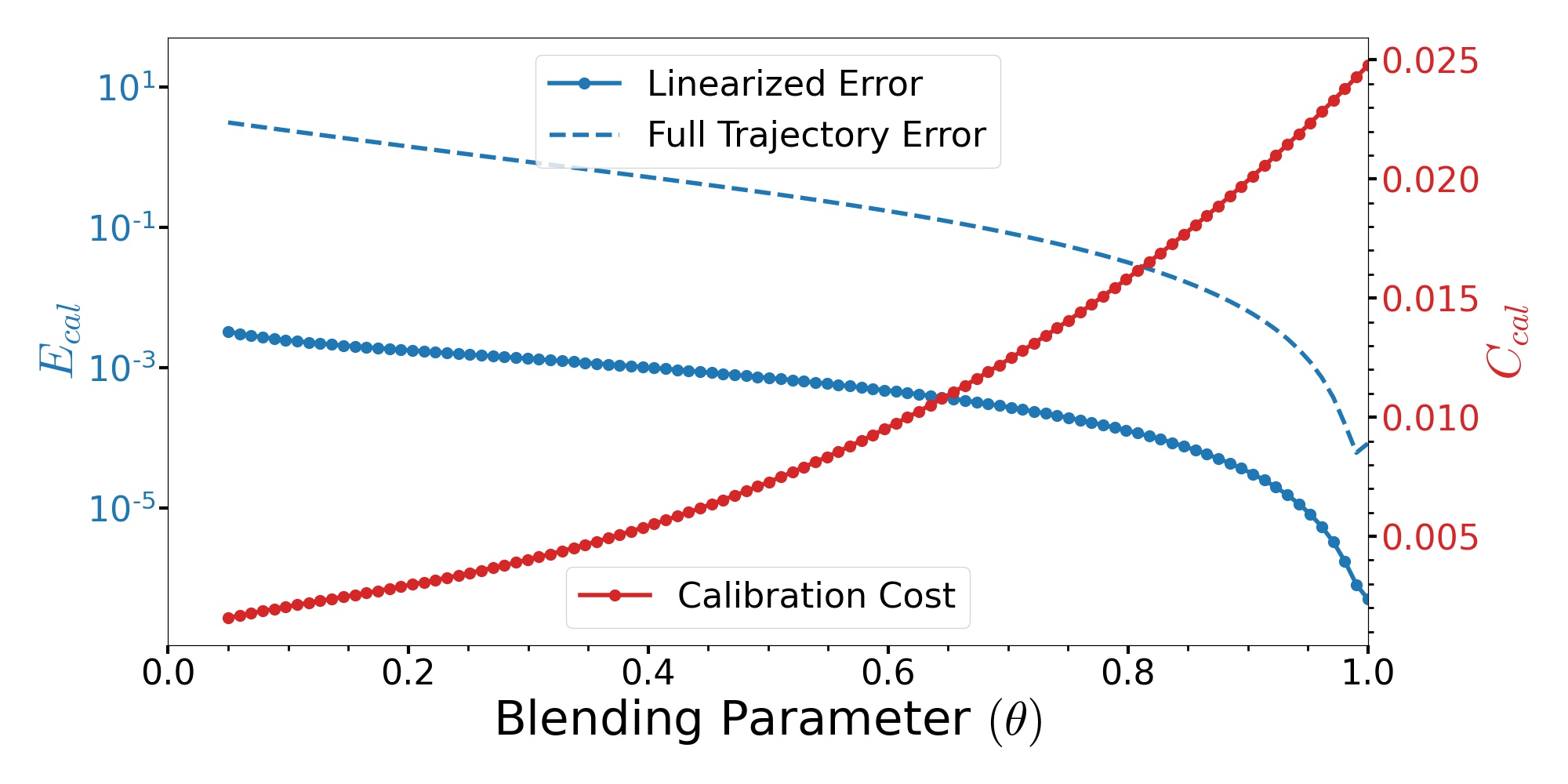}}
    \subfigure[$\theta$ close to 1($\theta \in (0.9,1)$)]{\includegraphics[width=0.36\linewidth]{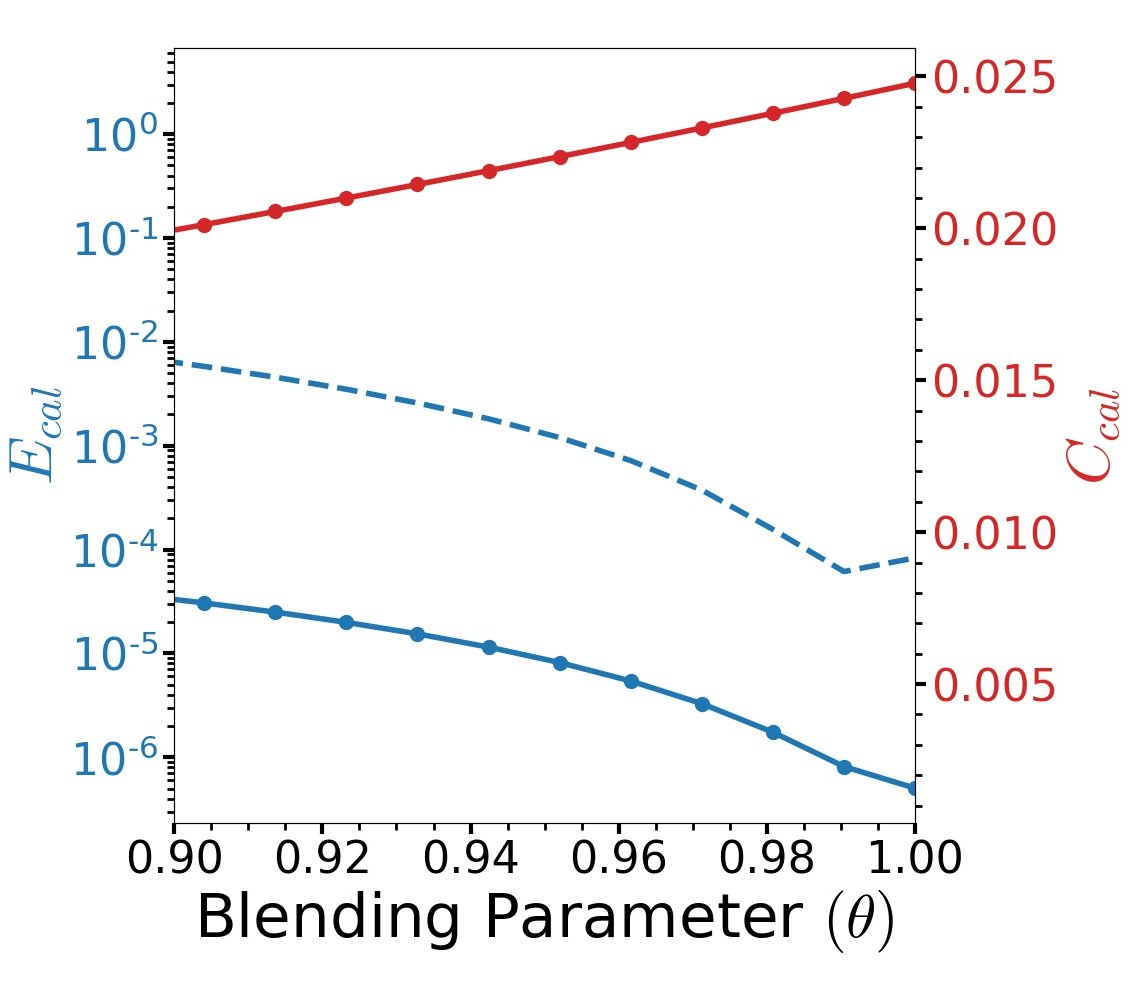}}
  \end{subfigmatrix}
  \caption{Variation of calibration cost~($\mathcal{C}$) and prediction error~($\mathcal{E}$) with blending parameter~($\theta$)}
  \label{fig:CalibThetaVar}  
\end{figure}

Figure~\ref{fig:CalibThetaVar} shows the variation of the calibration cost and linearized error as a function of the blending coefficient~($\theta$). 
The plot also shows the full trajectory error of the ROM calculated as the $L_2$ norm of the error between the reference POD time coefficients and those obtained from the ROM prediction over the entire calibration horizon similar to eq.~\ref{eq:lossDef}. 
Increasing values of $\theta$ represent a progressively diminishing relative importance of the calibration cost in the optimization problem and thus, result in a reduction of the linearized ROM prediction error and an increase in the calibration cost, as expected. 
However, for all values of $\theta$, the full trajectory error is several orders of magnitudes larger than the linearized prediction error. 
This is because the latter, by definition~(eq.~\ref{eq:err_def}), does not account for the cumulative nature of errors in predicted trajectories when integrating the ROM-ODE with inaccuracies in the coefficients. 
While both the linearized and full trajectory error exhibit a similar variation~(decreasing rapidly with increasing values of~$\theta$), this trend breaks down for $\theta$ values very close to $\theta=1$ where the full trajectory error increases even with a decrease in the linearized error metric, as shown in fig.~\ref{fig:CalibThetaVar}(b). 
This is likely the result of the near singular nature of the problem for values of $\theta$ close to one~(Appendix~\ref{sec:GPROMDetails}).
These variations of the error metrics and calibration costs during the calibration procedure are now compared to those obtained during the training process of the NeuralGP algorithm.

To ensure a illustrative evaluation, we initialized the values for the trainable ROM coefficients in the NeuralGP ROM to those obtained from the Galerkin Projection procedure. 
While this is unnecessary for the convergence of the DiffProg algorithm as shown by~\citet{mohan2021learning} who used random numbers to initialize the ROM-ODE coefficients, it affords key insights into the learning process by enabling a meaningful tracking of parameters similar to those used for the calibration method. 
To study this, the linearized and full trajectory error~(NeuralGP loss) as well as an equivalent calibration cost are tracked throughout the learning process and compared to those observed in the calibration procedure. 
Since the NeuralGP process does not consider a calibration cost for the error minimization, we define an equivalent calibration cost as: 
\begin{equation}
  \mathcal{C}_{learn} = \frac{||C^{learn}-C^{GP}||^2+||L^{learn}-L^{GP}||^2}{||C^{GP}||^2+||L^{GP}||^2}
  \label{eq:equivalentCalibCost}
\end{equation}
where $C^{learn}$, $L^{learn}$ and $C^{GP}$, $L^{GP}$ refer to the learned and Galerkin Projection based ROM-ODE coefficients respectively. 

\begin{figure}
  \centering
  \includegraphics[width=0.7\linewidth]{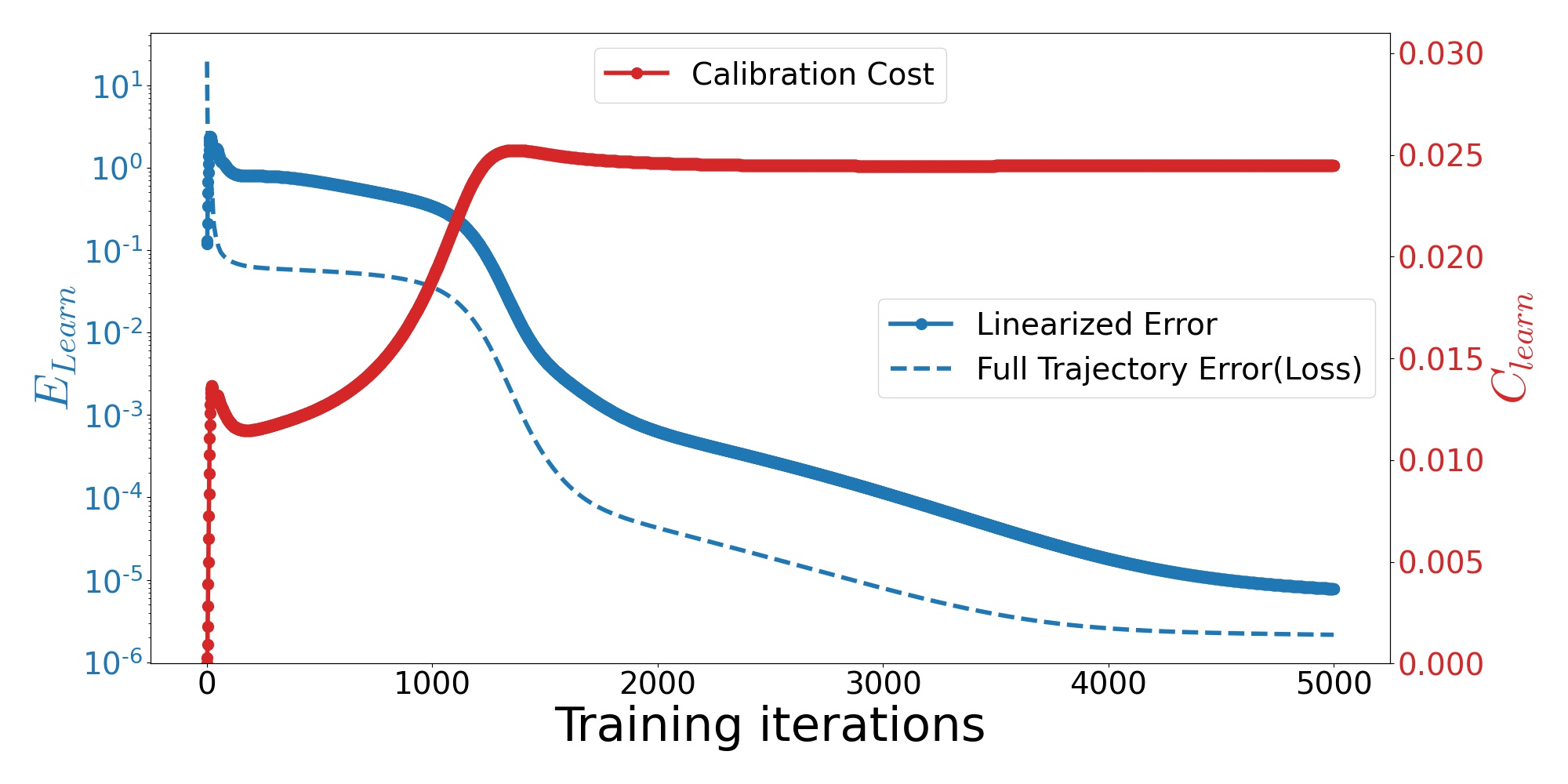}
  \caption{Variation in prediction error~($\mathcal{E}$) and equivalent calibration($\mathcal{C}$) costs with training iterations}
  \label{fig:LearnHist}
\end{figure}

Figure~\ref{fig:LearnHist} shows the variations in these performance metrics throughout the training process of the Neural GP procedure.
The full trajectory error exhibits a monotonic decrease over the course of the training, highlighting the ability of the DiffProg algorithm to converge rapidly to the trainable parameters of the actual system dynamics governing the ROM-ODE. At the end of the training, the full trajectory error obtained using the NeuralGP algorithm is \textit{lower} than that obtained in the calibrated-GP-ROM with even the highest values of $\theta$. 
This illustrates the superior performance of the NeuralGP algorithm in long-term predictions compared to those obtained from the calibrated GP-ROM.

Some additional important differences in the variation of the error metrics compared to their behavior for the calibration process~(shown in fig.~\ref{fig:CalibThetaVar}) can also be observed. 
Except for the initial few learning iterations, at all points during the learning process, the full trajectory error is lower than the linearized definition. 
Furthermore, the calibration cost grows initially to a value slightly greater than $0.025$ before marginally diminshing and remaining constant throughout the learning process. 
Even with a constant value of the calibration cost, however, both metrics of the prediction error decrease by several orders of magnitudes. 
This is an important difference compared to the calibration process, where the calibration cost increases monotonically with a decrease in the prediction error. This indicates that the Frobenius norm chosen for the calibration cost~(eq.~\ref{eq:calib_def}) is not fully representative of the actual impact of the calibration matrix on the system dynamics and may form a poor constraint on the calibration problem. 
Interestingly, even without considering calibration cost~\textit{i.e} with a $\theta$ value close to one, the NeuralGP procedure remains more efficient in reducing the achieved total trajectory error.

These trends highlight the two key advantages of NeuralGP technique:
the choice of the error metric being minimized and the consideration of a calibration cost.
Specifically, unlike the calibration procedure, the NeuralGP procedure seeks to minimize the \textit{full trajectory error} in prediction with no constraint on the calibration cost. 
Besides the above general trends, an interesting observation regarding the NeuralGP training procedure over the first few training iterations is that the full trajectory error decreases sharply with a rapid increase in the linearized error metric and the calibration cost. 
This is distinct from the general trend of the full trajectory and linearized error metrics having a similar qualitative variation~(decreasing or increasing together). 
This rapid change in the error metrics over the first few iterations is followed by a more gradual change where both error metrics exhibit a qualitatively similar decay while the calibration cost increases over the subsequent learning iterations. 
The rate of decay in both error metrics increases again at approximately thousand training iterations and continues to decay more quickly for the rest of the training epoch. 
We explore this behavior over the first few training iterations in a later section.

The differences in the trends exhibited by the ROM prediction metrics for the two methods are now explored further by considering their respective system dynamics. 
To derive a theoretical basis for comparison between the two ROM methodologies, their ROM-ODE coefficients are analyzed. 
First, the relative importance of the terms on the RHS of eq.~\ref{eq:ROM_ODE} were scrutinized over the short term prediction horizon by individually calculating their contribution to the temporal rate of change $\dot{a}$.
\begin{figure}
  \centering
  \includegraphics[width=\linewidth]{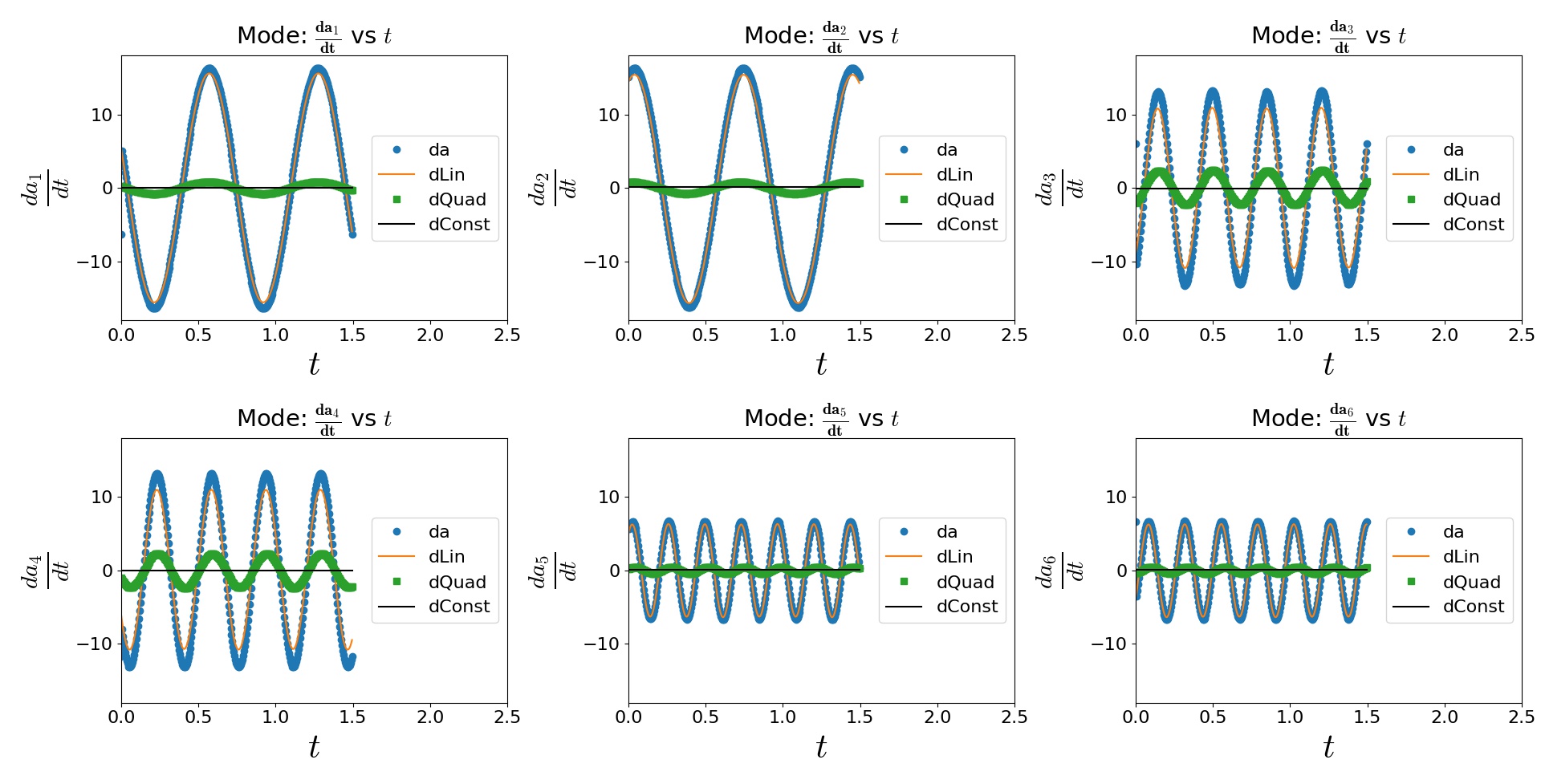}
  \caption{Relative magnitudes of the terms on the RHS of the ROM-ODE~(eq.~\ref{eq:ROM_ODE})}
  \label{fig:SysDynamics}
\end{figure}

Figure~\ref{fig:SysDynamics} compares the magnitudes of the change in the ROM coefficients at each time step~($da$) to the contributions from the constant~($dConst$), linear~($dLin$) and quadratic~($dQuad$) terms for the first six POD modes. 
The linear term has the largest impact on the system dynamics for all the leading modes tested, contributing to most of the rate of change at each time step. 
In comparison, the quadratic term has a contribution one order of magnitude smaller than that of the linear term while the impact of the constant term is smaller still. 
In view of the dominant influence of the linear term on the ROM-ODE~(eq.~\ref{eq:ROM_ODE}), we analyze the eigenvalues of the linear matrix~($L_{ij}$) to explain the impact of calibration and the NeuralGP method on the system dynamics.

\begin{figure}
  \begin{subfigmatrix}{2}
    \subfigure[NeuralGP]{\includegraphics[width=0.9\linewidth]{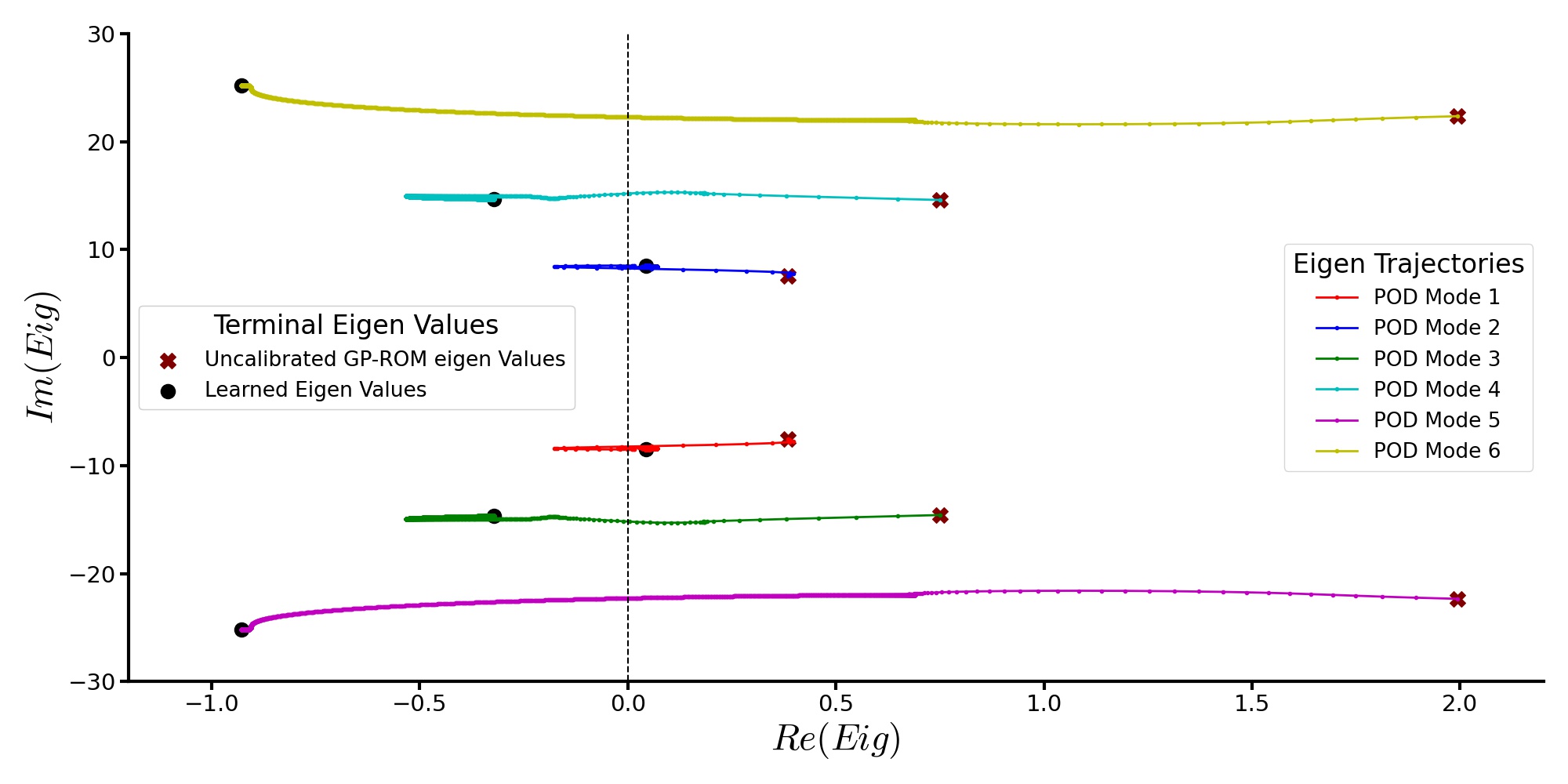}}
    \subfigure[Calibration]{\includegraphics[width=0.9\linewidth]{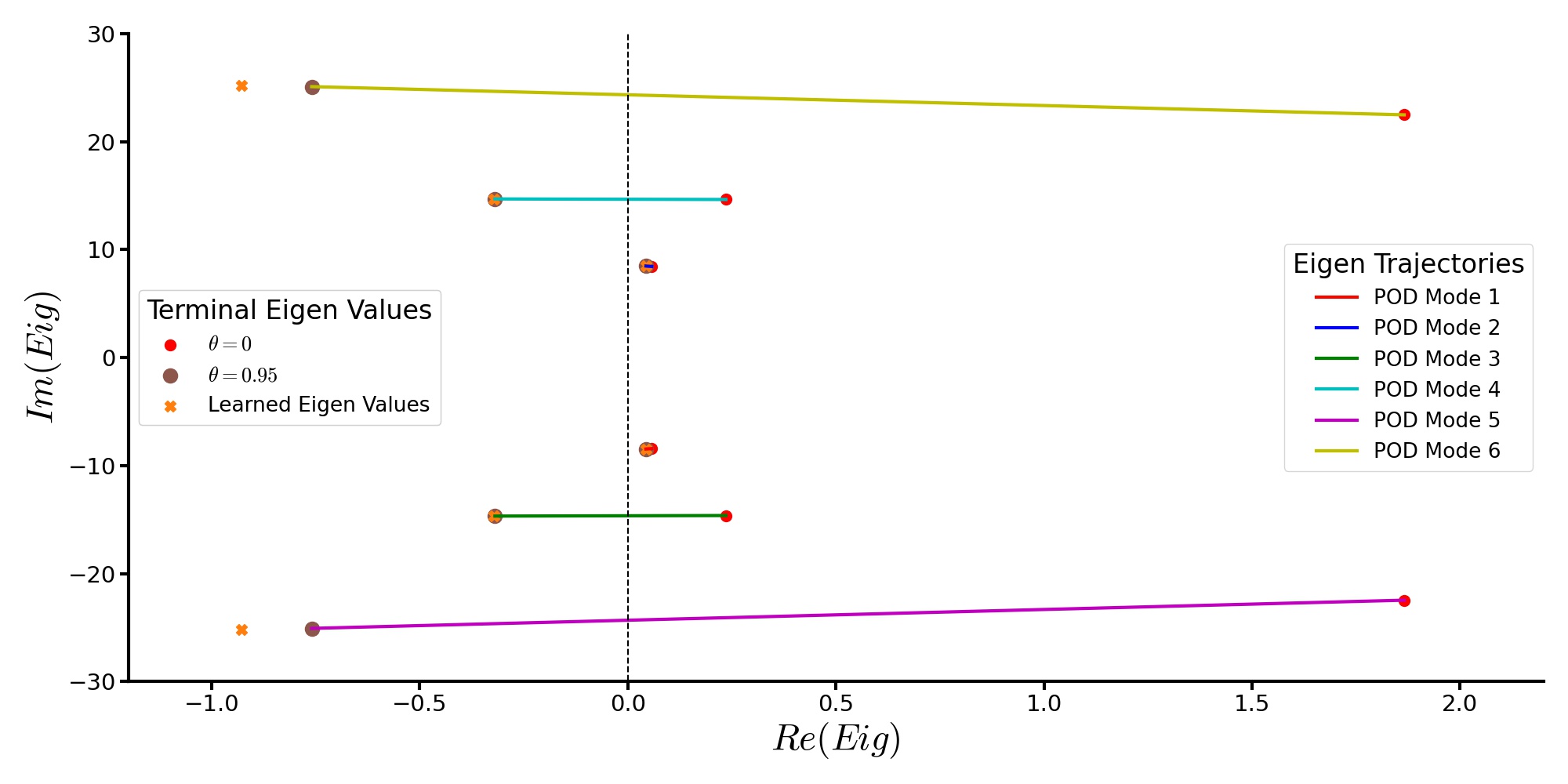}}
  \end{subfigmatrix}
  \caption{Trajectories of the eigenvalues for both the calibration and NeuralGP procedures.}
  \label{fig:EigenTrajectories}  
\end{figure}

The trajectories of the eigenvalues with a variation in $\theta$ and training iterations for the calibrated GP-ROM and the NeuralGP method respectively are shown in fig.~\ref{fig:EigenTrajectories}. 
Throughout their trajectories, the eigenvalues occur as conjugate pairs with the individual eigenvectors corresponding to the POD mode pairs as labeled. 
The perfectly conjugate eigenvalues reflect the fact that the leading POD mode pairs have the same temporal frequencies separated by a $90^o$ phase difference.
The $x$-axis $Re(Eig)$ in Fig~\ref{fig:EigenTrajectories} denotes the real part of the eigenvalues, and $Re(Eig) > 0$ in the positive half-plane indicates that the associated modes are unstable, i.e., growing with time. 
Similarly, $Re(Eig) < 0$ in the negative half-plane indicates modes that are decaying with time, while $Re(Eig) = 0$ are neutral modes that are neither growing nor decaying. 
The ROM comprises a linear combination of these modes, and their individual stability characteristics dictate the overall ROM stability. 
Therefore, the ROM would become inaccurate if all the modes are growing or decaying. 
A ROM typically comprises modes that have both $Re(Eig) < 0$ and  $Re(Eig) > 0$, and the challenge is to construct the modes with stability values that ensure the ROM does not diverge. 
The dominant modes also have a higher impact on ROM stability than the lower modes, as can be inferred from Fig.~\ref{fig:POD_energy}. 
As a result, both the calibration in GP-ROM and the differential programming approach in NeuralGP aim to perform a balancing act between mode dominance and its stability, to ensure ROM accuracy for as long as possible. 

The physical mechanism associated with the instability of the uncalibrated GP-ROM equations can be inferred from the corresponding eigenvalues of the uncalibrated $L_{ij}$ matrix.
We show this analysis in fig.~\ref{fig:EigenTrajectories} where cross-markers denote the initial eigenvalues of $L_{ij}$ before learning. 
These initial eigenvalues have positive real parts with progressively greater values for higher POD modes. 
The greatest of these, corresponding to the mode~$(5,6)$ pair, exhibit real parts as high as $\approx 2.0$. 
This is consistent with the rapid destabilization of the $(5,6)$ mode pair within the first few oscillation cycles in the uncalibrated GP-ROM predictions as shown in fig.~\ref{fig:ROM_results}.  

Both calibrated GP  and NeuralGP ROM stabilize the eigenvalues by ``damping" their real parts and making small corrections to their imaginary parts.
We plot these continuous changes to the eigenvalues with training as a trajectory in Fig.~\ref{fig:EigenTrajectories}(a), since it shows the intermediate eigenvalues obtained in the path to the final optimized $L_{ij}$. 
However, the trajectories followed for each ROM exhibit interesting differences. 
The NeuralGP method pushes the eigenvalues corresponding to the leading modes towards the negative half of the complex plane within a few initial training iterations~(every tenth training iteration shown with markers). 
Such rapid displacement of the eigenvalues quickly decreases the full trajectory error, and increases the equivalent calibration cost within the first few iterations of the training process, as seen in Fig.~\ref{fig:LearnHist}. 
Following this swift change in the leading eigenvalues during the initial training period, a reversal in their direction of motion is evident. 
A similar reversal is also observed for the (3,4) mode pair later during the training.
This is in contrast to the typical GP-ROM calibration process~(Fig.~\ref{fig:EigenTrajectories}(b)) which linearly moves the eigenvalues with only relatively small adjustments for the leading mode pair.

The explanation for the distinct trajectories of the eigenvalues lies in the differences in the optimization problem being solved in both cases. 
As discussed briefly in the previous section, the calibration process studied here~(based on Tikhonov regularization) solves a linearized form of a non-linear constrained optimization problem~\cite{cordier2010calibration}. 
Thus, a linearized error function~(eq.~\ref{eq:err_def}) is minimized during the calibration process and the movement of the eigenvalues also reflects the linearity of the optimization problem being solved with linear unidirectional trajectories.
In contrast, the NeuralGP method solves a non-linear minimization problem addressing the prediction error for the full trajectory across a prediction horizon. 
Since this algorithm integrates the ROM-ODE at each training iteration over a short time window, it captures the interactions between the modes. 
Thus, initially, the NeuralGP minimizes the full trajectory error by excessively damping the leading modes, while reducing the amplitudes of the highly unstable~$(5,6)$ mode pair as well.
This is a consequence of the coupled nature of the ROM-ODE wherein damping the leading modes also has the effect of damping the higher POD modes, due of their higher contribution to the RHS of the ROM equations. 
For example, the ROM equation for the time coefficient corresponding to the $a_5$ POD mode is given by:
\begin{equation}
\frac{d a_{5}}{dt} = C_5 + L_{5,1}a_{1} + L_{5,2}a_{2}+ L_{5,3}a_{3} + L_{5,4}a_{4}+ L_{5,5}a_{5} + L_{5,6}a_{6} + \sum_{j=1}^{N_{POD}} \sum_{k=1}^{N_{POD}} Q_{5,j,k}a_{j}a_k
  \label{eq:ModalCoupling}
\end{equation}
Here, the overpowering contribution of the leading $a_1$ and $a_2$ modes on the RHS is visible; reducing their magnitude has the effect of lowering the contribution of the higher $a_5$ and $a_6$ POD modes as well.
At the beginning of the training interval, the fastest method to reduce the prediction error is to damp the leading POD modes since correcting highly unstable eigenvalues of the $(5,6)$ mode pair requires a much more significant alteration of the dynamic matrix.
However, this trend is quickly reversed within a few training iterations when the inaccuracies generated by the additional dissipation added to the $(1,2)$ modes become a dominant factor in the total prediction error.

The position of the final stabilized eigenvalues in the complex plane also exhibit distinct and important features of the NeuralGP method. 
Initially, the most unstable eigenvalues corresponding to the $(5,6)$ mode pair are pushed far into the negative half-plane. 
Small corrections apply to the eigenvalues representing the leading POD modes~($1,2$ mode pair) whose stabilized eigenvalues retain slightly positive real parts. 
This shows that the two leading POD modes are, in fact, growing modes which are non-linearly saturated to yield a stationary solution.
Despite the overall similarities between the stabilized eigenvalues obtained using the two methods, a crucial difference emerges from fig.~\ref{fig:EigenTrajectories}(b). 
The NeuralGP procedure stabilizes the initially most unstable $(5,6)$ mode pair, shifting these values further into the negative half-plane. 
Even with a high value of $(\theta=0.95)$ for the blending parameter, the calibration procedure fails to stabilize this mode pair, resulting in discrepancies in the long-term predictions observed in fig.~\ref{fig:longTimeForecasts}. 
The large errors in long-term predictions from relatively insignificant changes in the eigenvalues serves to highlight the sensitivity of the ROM-ODE integration to its coefficients because of an accumulation of small errors at each time step.   

\section{Summary and Discussion}
\label{sec:conclusions}

A method using \textit{differentiable programming} is developed to construct a Machine Learning-based Reduced Order Model (ROM) that alleviates concerns about the black-box nature of the typical deep learning-based predictive tools while improving interpretability and prediction confidence.
The performance of the ROM is demonstrated and evaluated on a compressible, transonic flow over a NACA0012 airfoil, which contains shocks and vortices that can challenge the accuracy of long-time forecasting.
The results are compared with a traditional approach based on a Galerkin projection (GP) of the full Navier stokes equation on a truncated POD spatial basis, which is susceptible to instabilities as reported in the literature.

The GP-ROM calibration procedure required for accurate predictions using the uses optimization of linearized ROM prediction errors and incurs a calibration cost of a Tikhonov regularization framework. 
The differentiable programming method as used in the current method maintains the form of the ODE usually obtained in the GP-ROMs but learns the coefficients directly from training data using neural networks and gradient descent. Therefore, it effectively blends the strengths of both the ML and GP-based techniques for predictive modeling.  
The present approach displays several advantages over traditional black-box neural networks typically used in machine learning-based flow prediction tools~(LSTMs or CNNs).  
These include lower requirements for computational resources and training data while mitigating the risk of overfitting for a given data set. 
Although the short-time predictions are similar to GP-ROM methods, the new approach avoids the known divergence issues of calibrated GP-ROMs even for the longest prediction horizons tested..

A dynamical systems perspective of the ROM-ODEs provides insights into the distinct features of the proposed approach.
In particular, an examination of the eigenvalues identify the linear term in the ROM-ODE to be a dominant contributor to the system dynamics. 
The linear dynamics matrix from the uncalibrated GP-ROM-ODE contains eigenvalues on the right complex half-plane, indicating an exponential growth in the predictions, consistent with the observed instabilities of the ROM trajectories. 
Both the calibrated ROM and the current NeuralGP procedures stabilized the unstable eigenvalues by displacing them towards the left half of the complex plane with the final eigenvalues occupying similar positions, except for the higher POD modes, which are not adequately damped by the calibration procedure. 
Despite the general similarity between the stabilized eigenvalues, substantial differences become evident in the trajectories followed by the eigenvalues over the training/calibration process. 
These differences are shown to result from the NeuralGP algorithm effectively minimizing a more rigorous error norm.
The data driven nature, as well as lower error, bound achieved by the NeuralGP method, shows considerable promise in blending deep learning with physically interpretable methods and should facilitate the construction of ROMs for more complex flows, including fully turbulent flowfields.



\section*{Acknowledgements}
A.T.M and D.L. are supported by the LDRD (Laboratory Directed Research and Development) program at Los Alamos National Laboratory (LANL) under Contract No. 89233218CNA000001 with the U.S. Department of Energy/National Nuclear Security Administration under projects 20190058DR and 20220104DR. 

\appendix

\section{Details of the POD procedure}
\label{sec:podDetails}
The spatial POD basis functions~($\Phi_i=\{\Phi_i^{1/\rho},\Phi_i^u, \Phi_i^v,\Phi_i^w,\Phi_i^p\}^T$) are sought that maximize the projection of the fluctuation fields~($q(\mathbf{x,t})$) and optimally represent the second order statistics of the flow:
\begin{equation}
  \max_\mathbf{\Phi(\mathbf{x})} \frac{\overline{\langle q(\mathbf{x},t),\Phi(\mathbf{x})\rangle}_\Omega}{\langle \Phi(\mathbf{x}),\Phi(\mathbf{x})\rangle_\Omega}
  \label{eq:POD_defn}   
\end{equation}
Here, the $\overline{.}$ operator refers to a time averaging operation and the $\langle . \rangle_\Omega$ refers to the spatial inner product defined on the spatial domain $\Omega$, which, for two separate realizations of the state vector~($q^{I}$ and $q^{II}$), is given by:
\begin{equation}
  \langle q^{I},q^{II} \rangle_\Omega=\sum_{ivar=1}^{nvar}\frac{1}{\sigma_{ivar}^2}\int_\Omega q^{I}_{ivar}q^{II}_{ivar}d\mathbf{x}
  \label{eq:in_prod_def}   
\end{equation}
Thus, the contribution of the $i^{th}$ variable of the state vector~($q^{I}_{ivar}$) to the spatial inner product is normalized by  $\sigma_{ivar}^2$, the spatio-temporally averaged energy of the variable:
\begin{equation}
  \sigma_{ivar}^2=\int_\Omega \overline{q^2_{ivar}} d\mathbf{x}
\end{equation}
Using the above definition of the spatial inner product, the time cross correlation matrix is defined as:  
\begin{equation}
  K(t_1,t_2)=\langle q(\mathbf{x},t_1),q(\mathbf{x},t_2)\rangle_\Omega
  \label{eq:cross_corr_defn}
\end{equation}
Following the snapshot POD method~\cite{sirovich1987turbulence}, the maximization of the flowfield projections~(equation~\ref{eq:POD_defn}) is equivalent to solving the following eigenvalue problem to obtain the time variation of each spatial mode: 
\begin{equation}
  \overline{K(t_1,t)a_i(t)}=\lambda_i a_i(t_1)
\end{equation}
The above time coefficients~($a_i(t)$) are used to obtain the spatial basis functions by calculating the projection of the simulation snapshots on these time coefficients:
\begin{equation}
  \Phi_i(\mathbf{x})=\frac{1}{\sqrt{\lambda_i}}\overline{\tilde{q}(\mathbf{x},t)a_i(t)}
\end{equation}
Finally, the simulation snapshots~($q$) can be expressed using the time coefficients~($a_i(t)$) and the spatial basis functions~($\Phi_i(\mathbf{x})$) after a decomposition into mean~($\overline{q}$) and fluctuating component~($\tilde{q}$).
The optimality of the POD spatial basis then allows for an error bound truncation using only the leading $N_{POD}$ modes: 
\begin{equation}
  q=\overline{q}+\tilde{q}=\overline{q}+\sum_{i=0}^\infty a_i(t)\Phi_i(\mathbf{x}) \approx \overline{q}+\sum_{i=0}^{N_{POD}} a_i(t)\Phi_i(\mathbf{x})
  \label{eq:POD_decomp}
\end{equation}

\section{Details of the GP-ROM coefficients}
\label{sec:GPROMDetails}

The coefficients of the quadratic ROM-ODE~(eq.~\ref{eq:ROM_ODE}) are given by:
\begin{align}
  C_i&=\langle F^\alpha_{11}-A^\alpha_{11} ,\Phi_i\rangle_\Omega\\ \nonumber
  L_{ij}&=\langle F^\alpha_{1(j+1)}+F^\alpha_{(j+1)1}-A^\alpha_{1(j+1)}-A^\alpha_{(j+1)1} ,\Phi_i\rangle_\Omega\\
  Q_{ijk}&=\langle F^\alpha_{(j+1)(k+1)}-A^\alpha_{(j+1)(k+1)} ,\Phi_i\rangle_\Omega \nonumber
\end{align}
where the Greek indices ~($\alpha,\beta$) indicate a summation over the coordinate directions. The $F_{ij}$ and $A_{ij}$ matrices are given by:
\begin{equation}\label{eq:AF}
  A^i_{jk} = \begin{bmatrix}
    \Phi_j^{\star u_i} \Phi_{k,i}^{\star (1/\rho)} - \Phi_j^{\star (1/\rho)} \Phi_{k,i}^{\star u_i} \\
    \Phi_j^{\star u_i} \Phi_{k,i}^{\star u_1} + \Phi_j^{\star (1/\rho)} \Phi_{k,i}^{\star p}\delta_{1i} \\
    \Phi_j^{\star u_i} \Phi_{k,i}^{\star u_2} + \Phi_j^{\star (1/\rho)} \Phi_{k,i}^{\star p}\delta_{2i} \\
    \Phi_j^{\star u_i} \Phi_{k,i}^{\star u_3} + \Phi_j^{\star (1/\rho)} \Phi_{k,i}^{\star p}\delta_{3i} \\
    \gamma \Phi_j^{\star p} \Phi_{k,i}^{\star u_i} + \Phi_j^{\star u_i} \Phi_{k,i}^{\star p}
  \end{bmatrix}
  \text{ and }
  F^i_{jk} = \begin{bmatrix}
    0 \\
    \Phi_j^{\star (1/\rho)} \tau^\star_{1ik,i} \\
    \Phi_j^{\star (1/\rho)} \tau^\star_{2ik,i} \\
    \Phi_j^{\star (1/\rho)} \tau^\star_{3ik,i} \\
    \frac{\gamma \mu}{P_r} \left(\Phi_j^{\star p} \Phi_k^{\star (1/\rho)} \right)_{,ii} + (\gamma-1)\Phi_{j,i}^{\star u_\alpha} \tau^\star_{\alpha i k}
  \end{bmatrix}.
\end{equation}
Here, $\tau^\star_{ijk}=\mu(\Phi_{k,j}^{\star u_i}+\Phi_{k,i}^{\star u_j} - 2/3 \Phi_{k,\alpha}^{\star u_\alpha} \delta_{ij})$ and $\Phi^\star$ refers to the augmented POD basis comprising of the mean flow and the spatial POD modes: $\Phi^\star=[\overline{q}, \Phi_1,\cdots,\Phi_{N_{pod}}]$. The matrices $C_i$, $L_{ij}$, and $Q_{ijk}$ of equation~\ref{eq:ROM_ODE} can be pre-computed using only the spatial POD modes. The integration of the ROM equation itself requires $N_{POD}(1+N_{POD}+N_{POD}^2)$ scalar products per time step and is much cheaper than integrating the governing equations~(equation~\ref{eq:NS_equn}).

\bibliography{main.bib,Arvind_paper/jfm-instructions.bib}
\end{document}